\documentclass[sigconf,9pt]{acmart}
\AtBeginDocument{%
  \providecommand\BibTeX{{%
    Bib\TeX}}}

\setcopyright{cc}
\usepackage{microtype}
\usepackage{pifont}
\usepackage{algorithm}
\usepackage{algorithmicx}
\usepackage{algpseudocode}
\usepackage{amsmath}
\usepackage{graphicx}
\usepackage{booktabs} 
\usepackage{siunitx}  
\usepackage{multirow} 
\usepackage{hyperref}  
\usepackage[table]{xcolor}
\usepackage{multirow, makecell}
\usepackage{nccmath}
\usepackage{ragged2e}
\usepackage[caption=false,font=normalsize]{subfig}
\usepackage{threeparttable} 
\usepackage{xspace}
\acmYear{2026}
\copyrightyear{2026}
\setcctype[4.0]{by}
\acmConference[MobiSys '26]{The 24th Annual International Conference on Mobile Systems, Applications and Services}{June 21--25, 2026}{Cambridge, United Kingdom}
\acmBooktitle{The 24th Annual International Conference on Mobile Systems, Applications and Services (MobiSys '26), June 21--25, 2026, Cambridge, United Kingdom}
\acmDOI{10.1145/3745756.3809211}
\acmISBN{979-8-4007-2027-7/26/06}

\AtBeginDocument{%
  \providecommand\BibTeX{{%
    \normalfont B\kern-0.5em{\scshape i\kern-0.25em b}\kern-0.8em\TeX}}}

\newcommand{\ie}{\emph{i.e.},\xspace}

\newcommand{\eg}{\emph{e.g.},\xspace}

\newcommand\figref[1]{Fig.~\ref{#1}}

\newcommand\tabref[1]{Tab.~\ref{#1}}

\newcommand\secref[1]{Sec.~\ref{#1}}

\newcommand{\sysname}{{\sf AdaSprite}\xspace}

\ifodd 1

\else

\fi
\definecolor{mypurple}{RGB}{128,0,128}
\algrenewcommand{\algorithmiccomment}[1]{\textcolor{black}{/* #1 */}}

\begin{document}
\sloppy

\title{AdaSprite: Resource-efficient Online Co-Adaptation for V2I Systems Under Large-scale Data Drifts}

\author{Lehao Wang$^{\dagger}$, Zhiwen Yu$^{\dagger,+,*}$, Sicong Liu$^{\dagger,*}$, Kefan Chen$^{\dagger}$, Fengmin Wu$^{\dagger}$, Bin Guo$^{\dagger}$} 
\affiliation{%
  \institution{$^{\dagger}$Northwestern Polytechnical University, $^{+}$Harbin Engineering University}
  \country{}
}

\thanks{*Zhiwen yu and Sicong liu are corresponding authors.}


\begin{abstract}
The rise of vehicle–infrastructure (V2I) collaboration enables safer and broader perception. 
To process large-scale V2I video streams, vision-language models (VLMs) are promising as they unify multi-view vision into end-to-end task grounding, reducing handcrafted design.
We use Vision Mixture-of-Experts (V-MoE) as the distributed visual backbone of VLMs, leveraging sparse expert routing to enable \textit{conditional computation} across diverse viewpoints under resource constraints.
Yet, V-MoEs face a critical challenge: 
large-scale data shifts over minutes to hours in V2I systems, amplified by agnostic participants and biased features propagating through experts.
To maintain accuracy efficiently, we find it beneficial to co-adapt multiple V-MoEs on edge servers, avoiding the latency and privacy risks of cloud offloading and the accuracy sacrifices of on-device methods.
However, the resource-constrained edge poses challenges for efficient co-adaptation: i) DRAM fragmentation and imbalance limit expert parallelism, ii) memory-I/O bottlenecks restrict computation reuse, and iii) asynchronous adaptation increases task-switch overhead.
Also, prior work rarely explores the upper bound of concurrent tasks under limited edge resources, a critical factor for practical V2I deployment. 
To address these, we present \sysname. 
By combining cooperative elastic scaling with multi-level multiplexing, \sysname optimizes expert lifespans to reduce DRAM fragmentation, exploits predictable activation patterns for efficient I/O reuse, and employs twin-buffer scheduling to leverage sparsity.
On a weak edge, \sysname supports up to 17 concurrent V2I tasks (\textit{vs.} $\leq 6$ for baselines), improving SLO attainment by 1.6$\times$ and throughput by 2.1$\times$.
Also, it allows users to trade accuracy and concurrency for second-level adaptation.

\end{abstract}



\maketitle

\section{Introduction}
\label{sec:intro}

The rise of mobile and embedded vision has enabled AI-powered perception in vehicle systems~\cite{padmanabhan2023gemel,lu2022turbo,guo2018foggycache,guo2018potluck,xu2022v2x,wu2024adaflow}. 
Relying on a single vehicle’s sensors, however, leaves blind spots (\eg occluded pedestrians~\cite{zhang2023occlusion}), motivating networked vehicle-infrastructure (V2I) perception.
In V2I, particularly in the intermediate collaboration~\cite{chen2023transiff}, vehicles and roadside nodes \textit{share} and \textit{fuse} features to extend sensing range, avoiding raw-video backhaul and enabling tunable latency–quality trade-offs and enhance safety-critical tasks such as path planning~\cite{mao2023gpt}, localization~\cite{bai2023qwenvl}, and navigation~\cite{jain2022ground}.


Contrary to common belief, vision-language models (VLMs) can process large-scale video streams more resource-efficiently than traditional deep models by aligning visual inputs with task semantics, \eg “whether to yield,” especially in bandwidth- and compute-constrained V2I systems.
In our tests, VLM-based V2I reduces inference latency by 22.6\% and transmission by up to 84.9\% compared to CooperNaut~\cite{cui2022coopernaut}, while maintaining comparable accuracy.
Unlike traditional vision models that capture all raw features, VLMs extract task-relevant cues (\eg pedestrians, vehicles) and discard irrelevant data.
This \textit{task-driven} alignment yields compact embeddings, making VLMs ideal for resource- and bandwidth-limited V2I scenarios.
Moreover, VLMs enable instruction-driven tasks, such as high-level visual QA, directly supporting the detect–decide–act loop.

\begin{figure}[t]
    \centering
    \includegraphics[width=.46\textwidth]{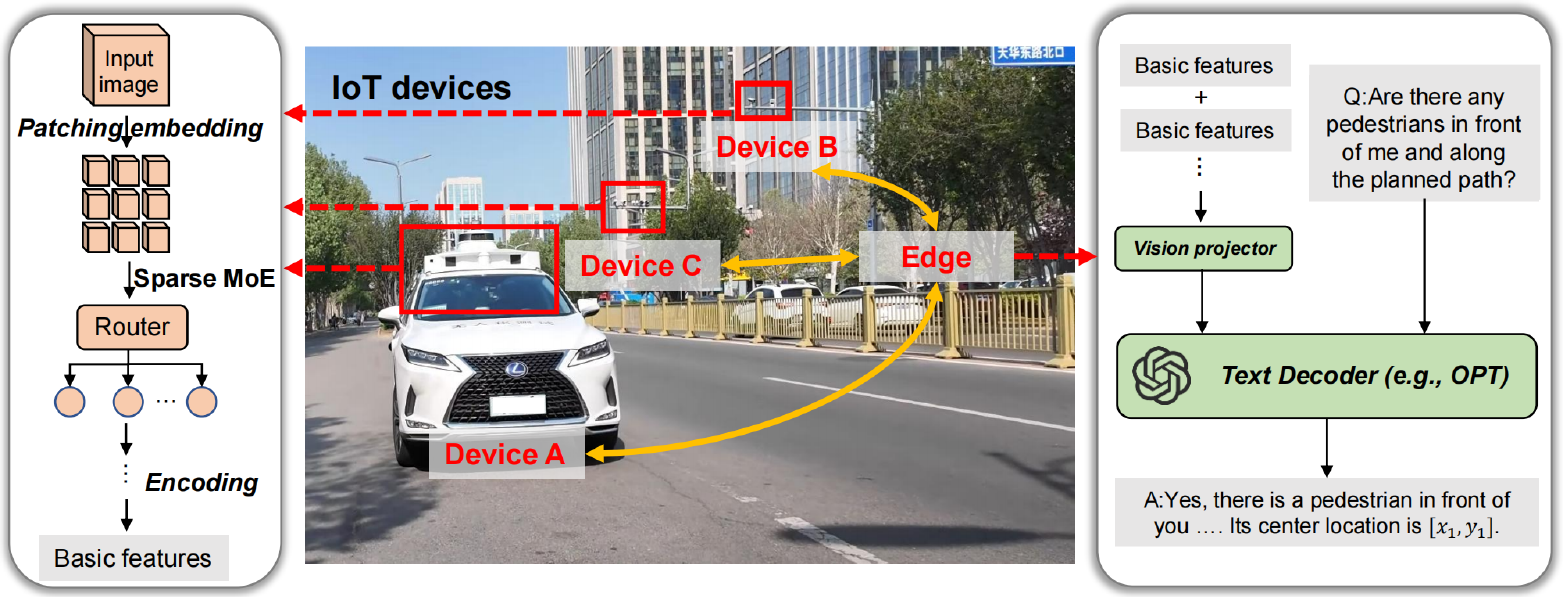}
    \vspace{-3mm}
    \caption{Showcasing networked V2I applications in urban traffic scenarios. 
    }
    \label{fig:moti_1}
    \vspace{-2mm}
\end{figure}

To further improve efficiency, we adopt Vision Mixture-of-Experts (V-MoE)~\cite{VMOE, chen2023mod} as a distributed feature extractor across heterogeneous V2I devices. 
Its \textit{sparse expert routing} and \textit{lightweight design} enable conditional computation, improving accuracy-efficiency trade-off while adapting to device-specific constraints.
As shown in \figref{fig:moti_1}, each device runs a compressed V-MoE tailored to its viewpoint: on-vehicle cameras (Device A) focus on dynamic obstacles with 15Mbps streams; roadside cameras (Device B) capture lateral context with 10Mbps streams; intersection cameras (Device C) track global traffic flow with 20Mbps streams. These object-aware embeddings are fused at the edge for downstream tasks, \eg planning and cooperative inference, enabling timely hazard warnings and improving traffic safety.

A key challenge is \emph{non-stationary distributed (large-scale) data drifts}. 
While the \textit{language} module of VLMs generalizes well, \textit{visual} modules deployed on distributed V2I devices continuously encounter long-horizon data drifts between training and real-world traffic, degrading accuracy. 
Agnostic participants (\eg vehicles) increase drift frequency, and shared V-MoE outputs propagate bias or errors (\textit{second-hand drift}). 
In \tabref{tab:accuracy-loss}, a drift in a single vision module can cause 35.73\% system-wide accuracy loss, which is unacceptable.

\emph{Adaptation} with targeted visual updates~\cite{khani2023recl,wang2023adaevo} can mitigate this. 
However, full \textit{cloud offloading}~\cite{kai2020collaborative} raises privacy and latency issues, while on-device tuning~\cite{huang2023elastictrainer,fang2024adashadow,niu2022efficient} sacrifices accuracy for efficiency. 
\textit{Edge-assisted adaptation} strikes a balance, asynchronously serving parallel jobs, sharing resources, and enabling a tunable accuracy-latency trade-off under continuous non-stationary drifts (\eg traffic-flow changes), as in Microsoft Edge Video Services~\cite{bhardwaj2022ekya, EVS}.
Also, prior work rarely explores the \textit{upper bound of concurrent adaptation tasks} under limited edge resources, a critical factor for practical V2I deployment. 
As participation scales, constrained GPU memory (DRAM) limits the number of tasks that can maintain accuracy with low latency. 
For example, a single RTX 3090 edge server can typically handle only six concurrent adaptation tasks using existing methods (see \secref{exp_concurrency}), with latency rising sharply as load increases.
For example, on a shared RTX 3090 edge server, only a 16\% increase in task load can cause a 40\% rise in average latency.

Our key insight is that \emph{co-adaptation}, enabled via \textit{cooperative elastic scaling} and \textit{multi-level reuse}, makes large-scale online adaptation feasible and effective (\secref{sec:challenge}).
\textit{First}, transformer-based V-MoEs have weaker inductive bias than CNNs; co-adaptation leverages parameter sharing and cross-task expert transfer to improve accuracy and increase the number of concurrent tasks. \tabref{tab:vertically-centered} shows co-adaptation reduces latency by 42.9\% compared to isolated adaptation while improving per-task accuracy.
\textit{Second}, vision workloads share substantial computation~\cite{shin2022algorithm}, creating redundancy. Isolated adaptation redundantly accesses identical DRAM-mapped regions, wasting memory and bandwidth. Co-adaptation instead uses a unified DRAM-resident expert cache, enabling \emph{no-copy, in-DRAM} reuse across tasks (Fig.~\ref{fig:moti_reuse}).

Realizing these benefits under \textit{dynamic-sparse} V-MoE adaptation with \textit{large-scale, non-stationary} mobile workloads raises three key challenges:

\noindent$\bullet$ \textbf{\textit{Challenge \#1: DRAM-constrained expert parallelism.}}
Although MoEs are parallelizable, inefficient DRAM usage caps co-adaptation parallelism, inflating latency and limiting concurrency. 
Two factors dominate:
\textit{i)} Dynamic, irregular DRAM requests from shared experts cause \textit{memory fragmentation}, reducing available DRAM for new tasks. 
Prior tuning or memory-efficient methods~\cite{wang2022melon,siebert2000eliminating,veldema2012parallel} improve footprint but often slow execution.
\textit{ii)} Sparse activations and heterogeneous experts, amplified by diverse V2I models, induce \textit{load imbalance}, causing stalls and high DRAM occupancy. 
Auxiliary balance losses~\cite{chen2023adamv} or strict capacity management~\cite{fedus2022switch} partially help but reduce accuracy due to suboptimal routing.

\noindent$\bullet$ \textbf{\textit{Challenge \#2: Memory-I/O-bound computation reuse.}}
Cross-task reuse can boost efficiency, but DRAM-constrained edge GPUs (\eg 10 GB on RTX 3080 vs. 21 GB reuse demand for V-MoE) require offloading redundant computation to disk, creating \textit{I/O bottlenecks}, especially with heterogeneous V2I streams. 
Dynamic sparsity and diverse models exacerbate irregular, bursty I/O, causing cache thrashing and latency spikes. 
Existing systems~\cite{huang2024freshgnn,guo2018foggycache} rely on frequent scattered I/O, while prefetchers~\cite{gao2024attentionstore} assume more regular access than MoE retraining exhibits. Eviction conflicts across shared experts further erode gains.

\noindent$\bullet$ \textbf{\textit{Challenge \#3: Scheduling asynchronous adaptation requests.}}
Non-stationary, asynchronous adaptation requests amplify dynamic sparsity, inflating task-switch overhead via repeated expert cache reloads. 
Heterogeneous accuracy gains across concurrent tasks complicate scheduling, making fairness-priority trade-offs critical to avoid inefficiency.

\sysname addresses these challenges with three mechanisms that turn limitations into opportunities.

\noindent$\bullet$ \textbf{\textit{First: align expert lifespans to reduce fragmentation and imbalance.}}
Long tasks can block shorter ones, so \sysname groups experts by lifespan and dependency using lightweight dynamic programming under a shortest-job-first principle~\cite{elmougy2017novel}. 
At runtime, resources are scheduled to align expert execution, reducing DRAM fragmentation and balancing load.

\begin{figure}[t]
    \centering
\includegraphics[width=.32\textwidth]{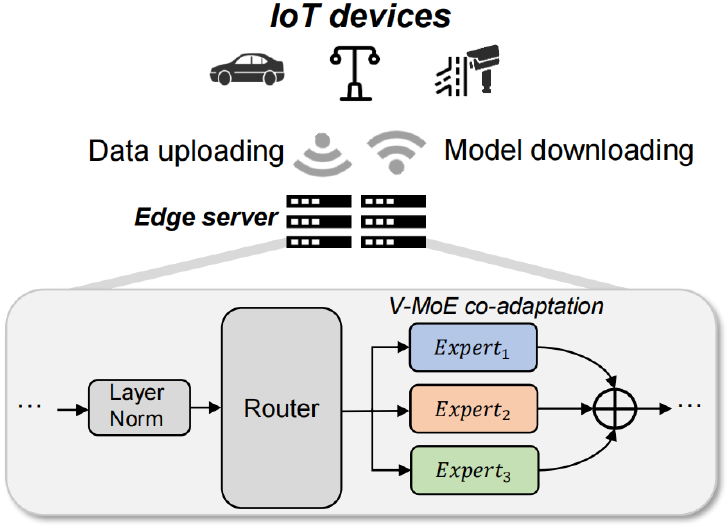}
\vspace{-1mm}
    \caption{Edge-assisted V-MoE co-adaptation.}
    \label{fig:moti_2}
    \vspace{-1mm}
\end{figure}

\begin{table}[t]
\centering
\sisetup{table-format=2.2}
\caption{Accuracy Loss from second-hand drift}
\vspace{-1mm}
\scalebox{0.75}{
\begin{tabular}{l *{6}{S}}
\toprule
\multicolumn{1}{c}{\textbf{Task}} & {\textbf{Exi.}} & {\textbf{Cou.}} & {\textbf{Obj.}} & {\textbf{Sta.}} & {\textbf{Com.}} & {\textbf{Overall}} \\
\midrule
\textbf{Accuracy loss (\%)} & 7.77 & 50.04 & 60.42 & 39.14 & 21.29 & 35.73 \\
\bottomrule
\end{tabular}
}
\label{tab:accuracy-loss}
\vspace{-1mm}
\end{table}

\begin{table}[tbp]
\centering
\sisetup{
  table-format=2.2,
  mode=text,
}
\caption{Comparing isolated- and co-adaptation}
\vspace{-1mm}
\scalebox{0.68}{ 
\begin{tabular}{
  l 
  S[table-format=2.2]
  S[table-format=1.2] 
  S[table-format=2.2]
  S[table-format=2.2]
  S[table-format=2.2]
  S[table-format=3.2] 
}
\toprule
\multirow{2}{*}{\textbf{Adaptation Method}} & \multicolumn{5}{c}{\textbf{Accuracy (\%)}} & \textbf{Avg. latency} \\
\cmidrule(lr){2-6}
& \multicolumn{1}{c}{\textbf{Exi.}} 
& \multicolumn{1}{c}{\textbf{Cou.}} 
& \multicolumn{1}{c}{\textbf{Obj.}} 
& \multicolumn{1}{c}{\textbf{Sta.}} 
& \multicolumn{1}{c}{\textbf{Com.}} 
& \multicolumn{1}{c}{\textbf{(s)}} \\
\midrule
Isolated adaptation & 49.68 & 8.62 & 15.17 & 34.74 & 52.05 & 385.25 \\
Co-adaptation       & 58.06 & 12.07 & 33.10 & 48.42 & 73.97 & 219.86 \\
\bottomrule
\end{tabular}}
\label{tab:vertically-centered}
\vspace{-1mm}
\end{table}

\begin{figure*}[t]
  \hfill
    \subfloat[In cloud] {\includegraphics[width=0.18\textwidth]{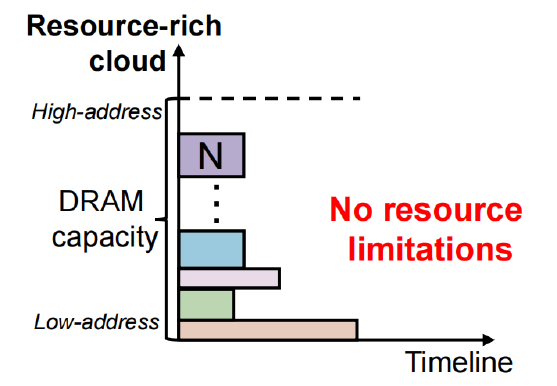}
    \label{fig:cloud}
    }
  \hfill
  \subfloat[At edge] {\includegraphics[width=0.18\textwidth]{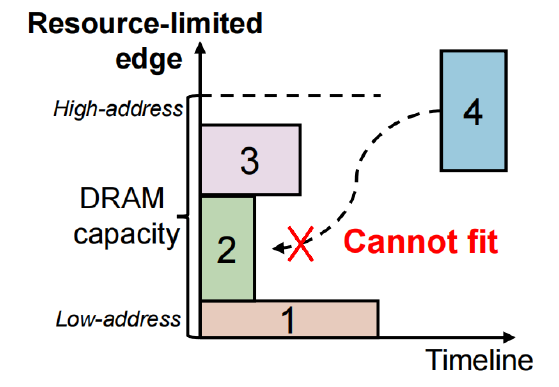}
  \label{fig:edge}
  }
  \hfill
    \subfloat[Comput. scale] {\includegraphics[width=0.18\textwidth]{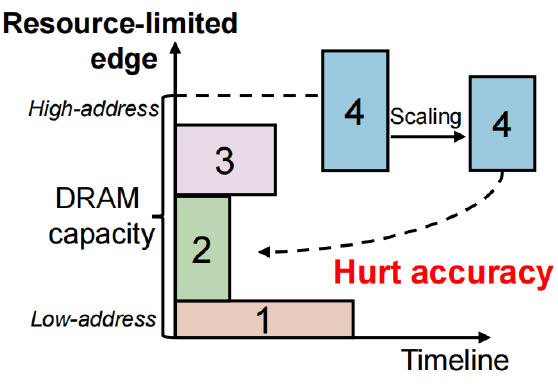}
  \label{fig:DRAM_1}
  }
  \hfill
    \subfloat[Comput. reorder] {\includegraphics[width=0.18\textwidth]{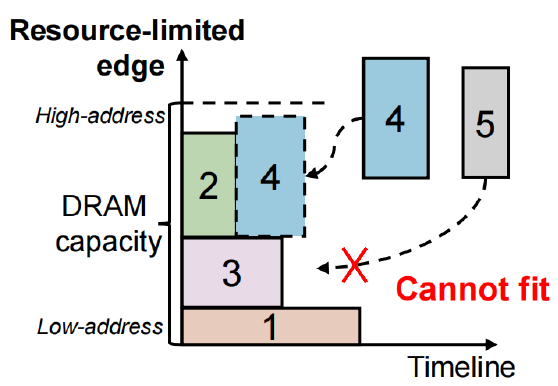}
  \label{fig:DRAM_2}
  }
  \hfill
    \subfloat[AdaSprite] {\includegraphics[width=0.18\textwidth]{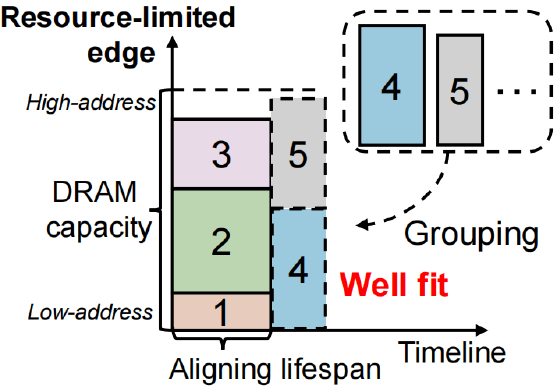}
  \label{fig:DRAM_3}
  }
  \hfill
    \vspace{-3mm}
  \caption{Comparison of various MoE parallelization strategies in resource-rich cloud or resource-limited edge.}
  \vspace{-2mm}
  \label{fig:DRAM}
 \end{figure*}

\noindent$\bullet$ \textbf{\textit{Second, exploit predictable activation trends for I/O-efficient reuse.}}
Although MoE retraining is dynamically sparse, the expert-activation probabilities are predictable. 
We design an \textit{I/O-hiding prefetch} that reorders input tokens so GPU compute overlaps with runtime I/O. 
An importance-aware token residency/eviction policy minimizes latency and mitigates cross-expert conflicts, converting dynamic sparsity into predictable, resource-efficient operation.

\noindent$\bullet$ \textbf{\textit{Third, harness sparsity with twin-buffer scheduling.}}
Rather than treating sparsity as a liability in adaptation, \sysname employs a \textit{twin-buffer} scheduler with two-tier GPU preemption~\cite{han2024pantheon}. It interleaves services across buffers and preloads \textit{inactive} experts during task switches, cutting reload overhead while balancing fairness and priority.

We implement \sysname as Linux \textit{microservices} (see \secref{sec:imple}).
We evaluate \sysname using three LLMs across five continuous traffic scenarios and five bandwidth settings on two self-collected continuous traffic datasets: \textit{CampReal} (a two-week real trace) and \textit{TrafficSim} (10+ road types, 150 vehicles, 38 cameras, and tens of concurrent task combinations). 
On a weak RTX 3090 edge server, \sysname sustains up to 17 concurrent adaptation tasks, whereas the best baseline supports at most 6 under the same configuration. \sysname also enables tunable concurrency–latency trade-offs; for example, reducing the task count to 4 yields second-level adaptation latency.
Overall, \sysname achieves the best accuracy–latency trade-off, cutting adaptation latency by up to 57.9\% and improving SLO attainment and throughput by up to 1.6$\times$ and 2.1$\times$, respectively. 
Even under extreme conditions (\eg expert-group failures), \sysname preserves significant gains, such as 42.7\% lower DRAM fragmentation.
Main contributions are summarized as follows.

\noindent$\bullet$
To our knowledge, this is the first work to examine the upper bound of concurrent adaptation tasks on weak edge servers. 
By treating expert \textit{dynamic sparsity} as an opportunity, \sysname enables cooperative elastic scaling and multi-level reuse for efficient MoE co-adaptation in V2I systems.

\noindent$\bullet$
\sysname introduces lifespan alignment, I/O-hiding prefetch with importance-aware residency/eviction, and twin-buffer scheduling, turning fragmentation, load imbalance, and asynchrony into throughput gains under non-stationary mobile scenarios.

\noindent$\bullet$
Across diverse V2I tasks/scenarios, \sysname consistently surpasses state-of-the-art in latency-accuracy trade-off.




\section{Background and Motivation}
\label{sec:motivation}

\subsection{The Trend of V-MoE in V2I Systems}
\label{sec:background}
Vision Mixture-of-Experts (V-MoE) augments Vision Transformers (ViTs) by replacing standard self-attention and feed-forward networks (FFNs) with expertized attention/MLP blocks, and by dynamically routing input tokens to the most relevant experts. 
V-MoE and its structure are increasingly adopted in VLM-based vehicle-infrastructure collaboration (V2I) for two reasons.
\textit{First}, ViTs typically serve as unified visual backbones in VLMs. V-MoEs preserve high-quality, consistent features across devices while avoiding task-specific backbone adaptation.
\textit{Second}, sparse expert activation enables conditional computation, improving accuracy-efficiency trade-offs by routing tokens to specialized experts.


\subsection{Primer of MoE Merging}
MoE merging treats experts from different V-MoEs as a parallel expert \textit{set} while merging shared components (\eg layer norms), enabling multiple V-MoEs to operate within a unified co-adaptation framework (~\figref{fig:moti_2}). 
This is well suited for V2I co-adaptation for two reasons.
\textit{i)} By keeping model-specific knowledge isolated within experts and merging only shared structures, V2I devices can contribute diverse experts to a common pool, improving concurrency through parameter sharing while avoiding interference and catastrophic forgetting.
\textit{ii)} Sparse activation yields minimal inter-expert dependency, allowing device-specific V-MoEs to be attached or detached \textit{asynchronously} from the co-adaptation pool with negligible accuracy loss~\cite{chen2023mod}.

\subsection{Challenges and Opportunities}
\label{sec:challenge}
In large-scale drifts, co-adapting experts not only improves model-level efficiency and knowledge fusion, but also \emph{breaks resource isolation}, enabling cross-model sharing of DRAM, compute, and I/O.
Under this shared context, challenges such as \textit{dynamic sparsity} and \textit{I/O constraints} can be reframed as opportunities for adaptation efficiency.

\begin{figure}[t]
\centering
\begin{minipage}[t]{0.225\textwidth}
\centering
\includegraphics[width=2.8cm]{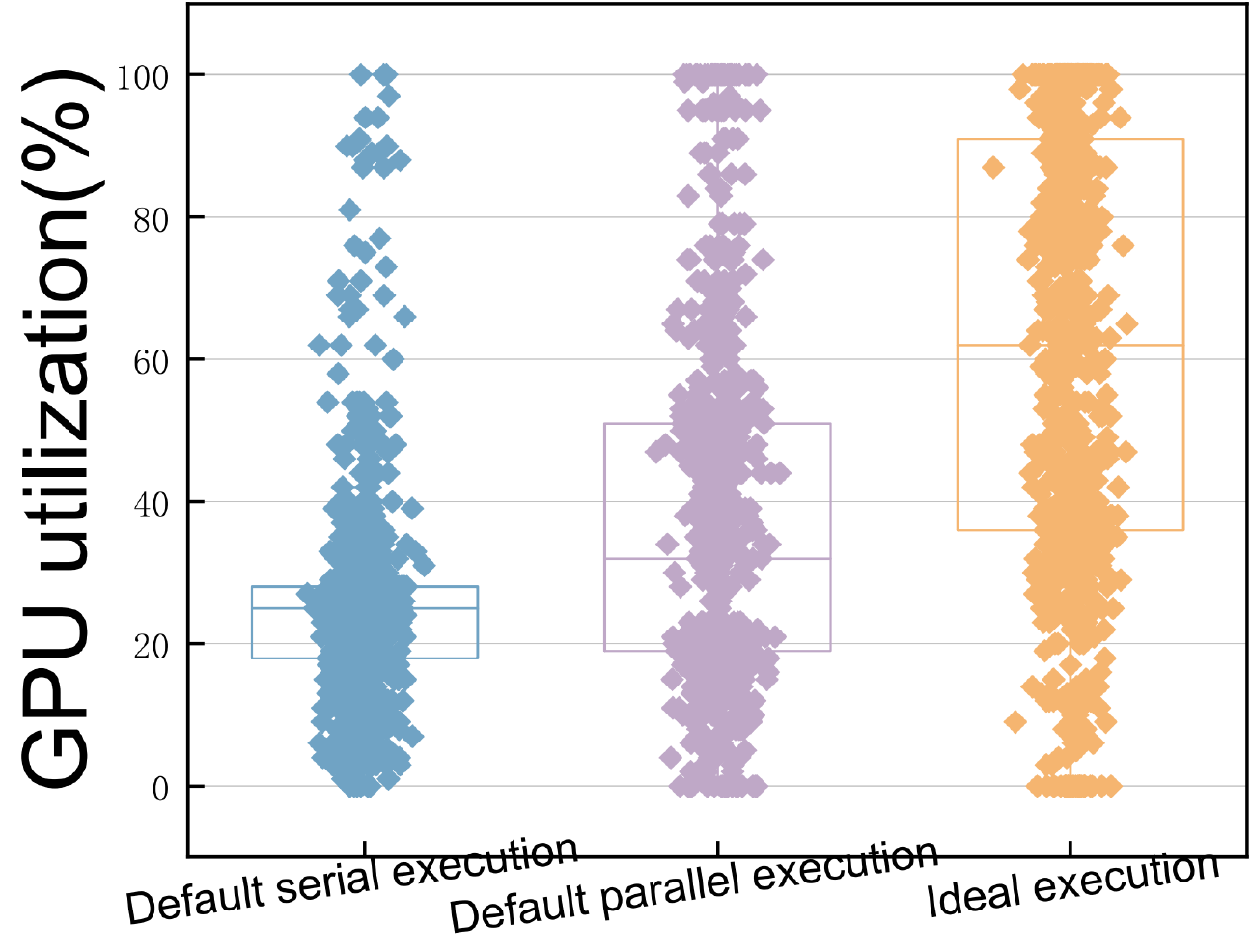}
\vspace{-3mm}
		\caption{GPU utilization during executions.}
		\label{fig:moti_utilize}
\end{minipage}
\hspace{0.01\textwidth} 
\begin{minipage}[t]{0.215\textwidth}
\centering
\includegraphics[width=3.4cm]{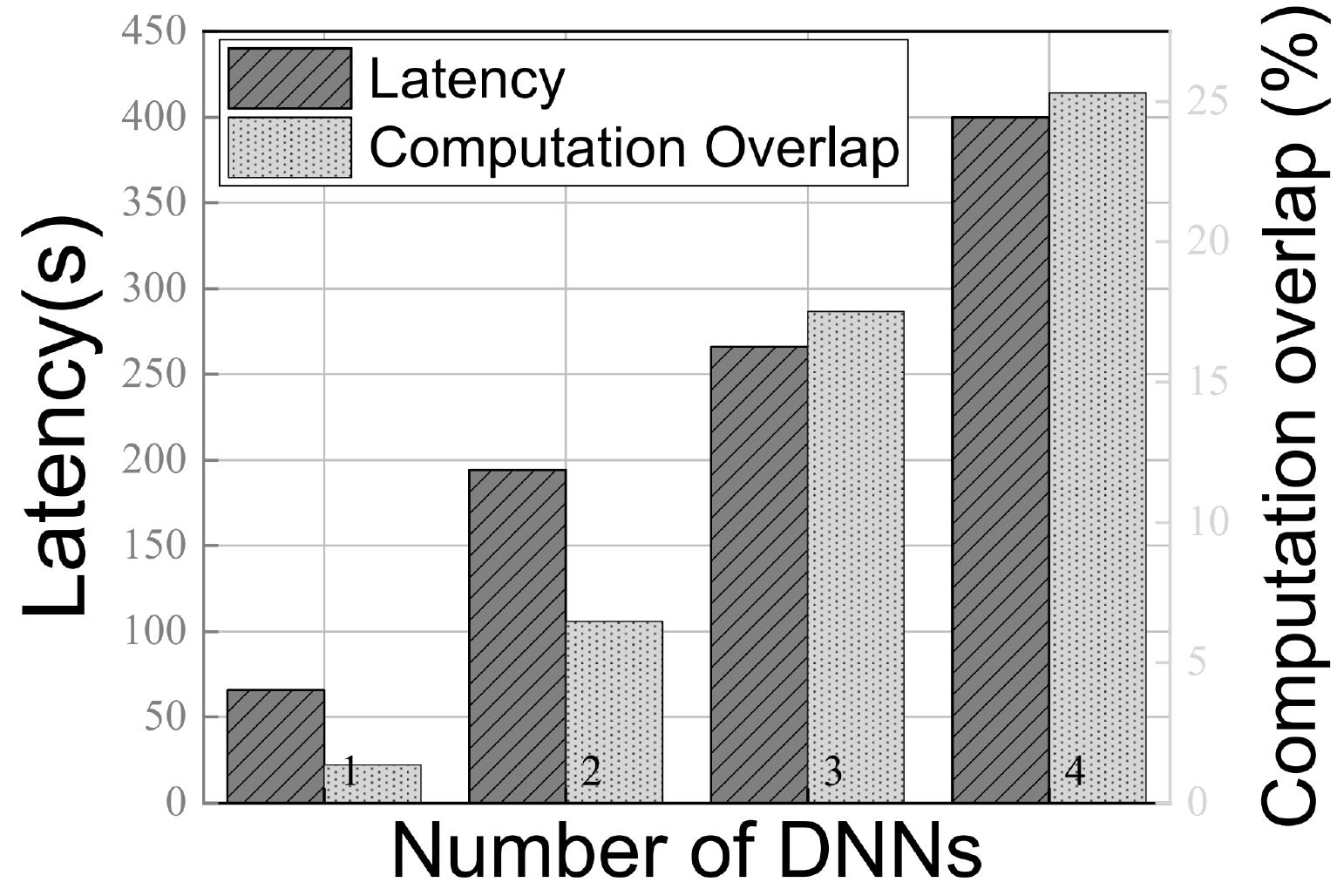}
\vspace{-3mm}
		\caption{Latency and computations.}
		\label{fig:moti_reuse}
\end{minipage}
\vspace{-2mm}
\end{figure}

\begin{figure*}[t]
    \centering
    \includegraphics[width=.67\textwidth]{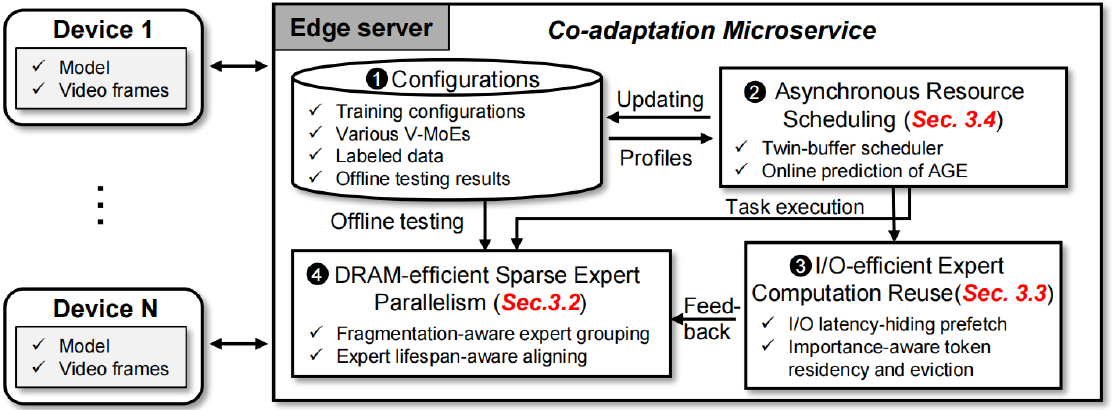}
    \vspace{-1mm}
    \caption{
    Facing new-arrival tasks with \ding{202}, \ding{203} determines adaptation tasks to service based on task demands and system states while updating \ding{202} through online profiling.
    During task execution, for each layer, \ding{204} identifies tokens for computation and feeds it back to \ding{205} which schedules the execution of experts to maximize their computation parallelism. 
    Meanwhile, \ding{204} reuses other tokens with low I/O overhead to improve I/O efficiency.
    }
    \label{fig:overall}
\end{figure*}

\subsubsection{Sparse Activation of the V-MoE}
Heterogeneous V2I devices trigger \emph{irregular, sparse} expert activations, which create two key bottlenecks at the edge.
\textit{First, fragmentation}.
Dynamic sparsity produces transient intermediates and irregular DRAM requests, forming small, discontinuous memory blocks that restrict parallelism and cap concurrency.
While cloud settings hide this issue with abundant memory (\figref{fig:cloud}), edge GPUs (\figref{fig:edge}) can host only a subset of experts per cycle, magnifying fragmentation.
Existing memory optimizations, scaling/repartitioning~\cite{nie2022tsplit}, compute reordering~\cite{wang2022melon}, swapping/recompute~\cite{ren2021zero,peng2020capuchin}, or fragmentation-oriented GC/compaction~\cite{siebert2000eliminating,veldema2012parallel,siegwart2006improving,hudson2001sapphire}, either incur accuracy loss or impose non-trivial overhead on weak edge servers.
\textit{Second, load imbalance}.
Uneven expert activations cause severe imbalance and synchronization stalls; measured GPU utilization under default parallel execution remains around 20\%, nearly collapsing to serial execution (\figref{fig:moti_utilize}).
Balancing losses~\cite{chen2023adamv} and capacity limits~\cite{fedus2022switch} mitigate load skew but degrade accuracy; cloud-optimized schedulers~\cite{nie2023flexmoe} expect abundant DRAM and interfere with other experts on the edge.

\textit{Asynchrony further worsens both issues}.
V2I devices upload models at different times, expanding the expert pool and diverging per-task activation sets. This leads to frequent cache reloads, high task-switch overhead, and heterogeneous retraining speeds.
Schedulers designed for isolated objectives~\cite{khani2023recl,bhardwaj2022ekya,gu2021liquid} cannot prevent reload overhead or capture time-varying, multi-task demands without expensive tuning.

\subsubsection{Redundant Computations and I/O Overhead}
Irregular MoE \textit{dynamic sparsity} in V2I makes computation reuse inherently \emph{I/O-bound}.
The working set of reusable activations (token features and expert outputs) frequently exceeds edge-GPU DRAM, forcing spills to external storage when concurrency grows.
For instance, adapting five V-MoEs with batch size 500 generates $>$20 GB reusable activations per epoch, far beyond 10 GB DRAM of an RTX 3080, resulting in 12 GB disk traffic and $\geq$25s I/O latency, nearly cancelling the 35 s compute saved by reuse.

Dynamic sparsity further produces \emph{irregular, short-burst access patterns} that thrash caches and increase latency (\figref{fig:baseline1}).
Existing fast-lookup~\cite{guo2018potluck}, reuse–accuracy trade-off~\cite{guo2018foggycache}, and selective-caching systems~\cite{huang2024freshgnn} reduce redundant compute but rely on frequent external I/O or assume regular access, incompatible with expert-driven sparsity.

\subsubsection{Turning Challenges into Opportunities}
V-MoE \emph{co-adaptation} breaks resource and knowledge isolation, revealing three opportunities that turn dynamic sparsity into a coordination signal rather than a competition/bottleneck.
\textit{i) Cross-model resource multiplexing.}
Pooling experts across models amortizes kernel and memory-management overheads.
Co-scheduling experts with complementary lifespans and dependencies reduces DRAM fragmentation and thread-block idling, improving parallelism without sacrificing accuracy.
\textit{ii) Unified, DRAM-resident reuse cache.}
A shared cache enables global prefetching and eviction decisions across models.
Token reordering provides \emph{I/O-hiding prefetch} that overlaps GPU compute with storage access, while \emph{importance-aware residency/eviction} preserves high-value experts and avoids eviction conflicts.
This keeps reuse effective even when working sets exceed device memory and suppresses I/O-induced latency spikes.
\textit{iii) Flexible task interleaving without per-model context.}
Removing strict per-model context enables lightweight task switches: we exploit natural asynchrony in V2I system to interleave tasks while preloading inactive experts, cutting reload overhead and easing fairness-priority trade-offs.
We realize them in \sysname via three mechanisms detailed next.

\section{System Design}
\label{sec:design}

To address the challenges and integrate the opportunities, \sysname introduces a V-MoE \emph{co-adaptation} service at the V2I edge. 
It coordinates concurrent adaptation jobs at both the computation and resource levels, meeting per-model accuracy-latency demands for open-world V2I applications.

\subsection{Overview} 
As shown in \figref{fig:overall}, \sysname separates the \emph{inference phase} from the \emph{adaptation phase}. 
During inference, mobile/IoT devices (on-vehicle, roadside, intersection) run local lightweight V-MoEs and transmit compressed features to the edge for multi-view fusion and downstream inference (vision projector + LLM decoder)
(\secref{sec:background}).
During adaptation, mobile/IoT devices sample and upload frames (and metadata) to the edge on demand~\cite{wang2023adaevo}, where a full-capacity reference ViT produces supervision (features or labels). 
\sysname packages each model’s data and objectives into an adaptation job, merges the corresponding V-MoEs into a unified co-adaptation pool via MoE merging, and manages jobs in isolation-at-the-data but sharing-at-the-resource fashion for fine-grained monitoring and control.
We follow the specification-based tuning method, \ie adapting only the visual module and projector to ensure efficiency.

\sysname optimizes co-adaptation via three core mechanisms that directly map to Challenges \#1-\#3 for large-scale data drifts:
\textit{(i) DRAM-efficient sparse-expert parallelism (\secref{sec:design_1}).}
Targeting \textbf{Challenge \#1} (fragmentation and load imbalance), \sysname employs lightweight dynamic programming to \emph{align expert lifespans} across tasks and performs spatio-temporal co-scheduling with dynamic resource allocation. This shrinks DRAM fragmentation, smooths parallelism, and improves utilization without sacrificing accuracy.
\textit{(ii) I/O-efficient expert computation reuse (\secref{sec:design_2}).}
Addressing \textbf{Challenge \#2} (I/O-bound reuse), \sysname enables a shared DRAM-resident expert cache for \emph{in-DRAM reuse} across tasks, and uses \emph{I/O-hiding prefetch} to overlap GPU compute with storage access. An \emph{importance-aware residency/eviction} policy prioritizes high-leverage experts, mitigating conflicts and minimizing external I/O even when the working set exceeds device memory.
\textit{(iii) Asynchrony-aware resource scheduling (\secref{sec:design_3}).}
For \textbf{Challenge \#3} (asynchronous requests), \sysname adopts a \emph{twin-buffer scheduler} with cross-task preemption: it interleaves services across buffers and \emph{preloads inactive experts} during task switches, cutting reload overhead while balancing fairness and priority.


\begin{figure}[t]
    \centering
    \includegraphics[width=.39\textwidth]{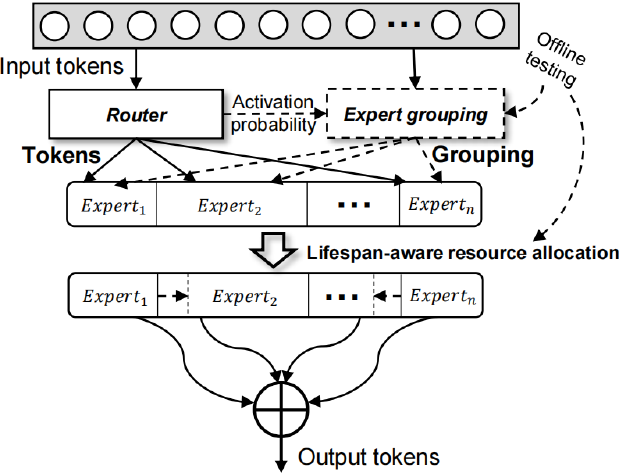}
    \caption{DRAM-efficient expert parallelism.}
    \label{fig:design1}
\end{figure}

\subsection{DRAM-efficient Expert Parallelism}
\label{sec:design_1}
Expert parallelism can accelerate V-MoE retraining, but on resource-limited edge GPUs it is constrained by \emph{DRAM fragmentation} (from dynamic sparsity) and \emph{barrier stalls} (from load imbalance), as discussed in \secref{sec:challenge}.

\textbf{Limits of Prior Reactive Methods.}
Previous MoE training systems~\cite{nie2023flexmoe} primarily focus on pretraining or single-model training to improve the throughput of individual models, while overlooking the efficient utilization of scarce resources.
And existing memory optimization~\cite{veldema2012parallel,siegwart2006improving} are fundamentally \emph{reactive}.
They intervene only \emph{after} fragmentation or imbalance appears, adding unpredictable overhead and implicitly assuming expert activations are uncontrollable randomness.
These arise from two structural mismatches:
\textit{i) Lifespan \textit{vs.} allocation.} Transient intermediates from sparse experts occupy low addresses and retire early; later allocations rarely fit these holes, accumulating fragmentation.
Heuristics (\eg high-address placement~\cite{wang2022melon}, scaled computations~\cite{nie2022tsplit}) partially help but risk accuracy or leave residual holes.
\textit{ii) Resource \textit{vs.} demand.} Uniform slicing ignores heterogeneous expert costs, causing heavy experts to dominate slice completion time and inflate synchronization stalls.

\textbf{Key Idea}.
Instead of reacting to symptoms, \sysname leverages shared context in modular/partially loaded components to \emph{proactively} restructure execution.
%
The complex mismatch collapses into two controllable dimensions: \textit{where} experts co-execute (space) and \textit{when} they complete (timing).

\noindent$\bullet$ \emph{Spatial grouping (lifespan alignment).} Experts pooled across models are partitioned into ordered \emph{time slices} by grouping complementary live ranges and dependencies.
This pre-allocates contiguous slots within a runtime-managed cache region and prevents irregular allocations from creating holes.

\noindent$\bullet$ \emph{Temporal alignment (balance-aware co-scheduling).} Within each slice, \sysname equalizes expert completion times via resource-aware scheduling, shrinking barrier stalls and smoothing parallelism under DRAM/GPU constraints.

As shown in \figref{fig:design1}, spatial grouping (\secref{sec:grouping}) removes most fragmentation \emph{before it forms}, while temporal alignment (\secref{sec:resource_all}) eliminates synchronization bottlenecks during parallel execution.

\begin{figure*}[]
\centering
\subfloat[Reuse without I/O optimization.]{
    \includegraphics[width=0.31\textwidth]{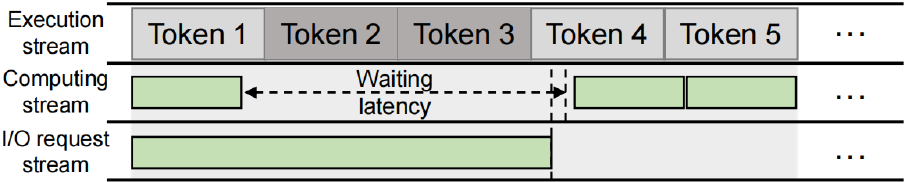}
    \label{fig:baseline1}
}
\subfloat[Reuse with prefetch.]{
    \includegraphics[width=0.31\textwidth]{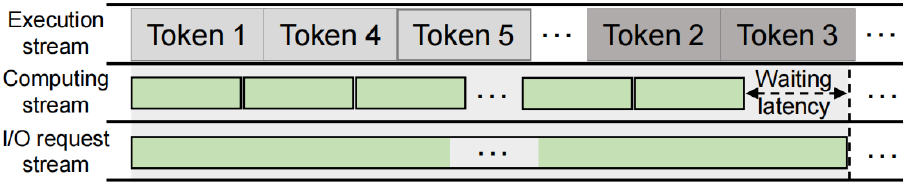}
    \label{fig:baseline2}
}
\subfloat[Reuse with prefetch \& residency.]{
    \includegraphics[width=0.31\textwidth]{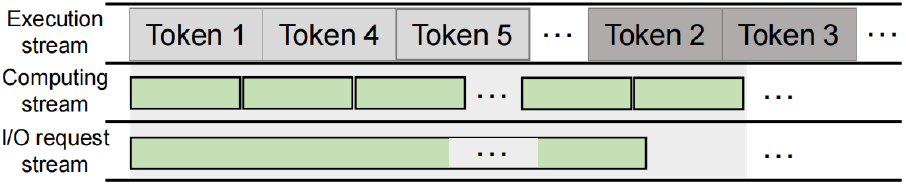}
    \label{fig:our_reuse}
}
\caption{Comparison of computation reuse methods.}
\label{fig:reuse}
\end{figure*}

\subsubsection{Fragmentation-aware Expert Grouping}
\label{sec:grouping}
Spatial grouping must assemble experts that \textit{tightly fit} DRAM and \textit{share compatible lifespans} to avoid stragglers that stall synchronization.
Random grouping leads to poor DRAM utilization and imbalance~\cite{mansi2024characterizing}, while aligning experts with large workload gaps extends the wall-clock time of lighter ones. Finding temporally optimal groupings under token-dependent routing is NP-hard, and existing compiler/graph schedulers~\cite{ma2020rammer,zhang2023cocktailer,kwon2020nimble,ding2021ios} prioritize compute throughput but ignore edge DRAM constraints.
To address this, \sysname employs fragmentation-aware expert grouping that combines fast per-expert DRAM estimation with a lightweight DP-based cross-task grouper, enabling efficient solutions to this coupled multi-objective problem.

\textbf{Fast per-expert DRAM estimator.}
We adopt a lightweight, \emph{lifespan-aware} DRAM usage estimator to support online grouping. 
In addition to well-researched static DRAM $M_{s}$~\cite{wang2023adaevo}, for expert $e$ with token count $T=b\cdot n_t$ (batch $b$, tokens per sample $n_t$) and key dimension $d_k$ in a ViT-style attention block, the temporary footprint decomposes into $M_{\mathrm{tmp}}(e)=m_c(e)+m_f(e),$
where $m_c(e)$ captures compute-dependent buffers (\eg Q \slash K \slash V projections and attention scores) and $m_f(e)$ aggregates function-call overheads (kernels, framework metadata) that are hardware\slash system-stable and can be profiled offline. Concretely,
$m_c(e)\approx B_p\big(\theta_1\,T\,d_k + \theta_2\,T^2\big),$
with $B_p$ the numeric precision (8\slash 1\slash 32-bit) and $\theta_1,\theta_2$ constants reflecting the implementation (queries\slash keys\slash values vs.\ score matrices). This estimator avoids irregular online tracing while preserving the key $T$- and $d_k$-scalings that dominate memory pressure.

\textbf{Cross-task expert grouping.}
We denote the estimated runtime (lifespan) as $r_e$, predicted in \secref{sec:resource_all}. 
Given a candidate set $\mathcal{E}$ of experts eligible to co-run and 
server DRAM budget $M_c$, select a slice $S\subseteq \mathcal{E}$ that (a) respects $\sum_{e\in S}M_{\mathrm{tmp}}(e)+ M_{s} \le M_c$, (b) \emph{maximizes} DRAM utilization $U_{\mathrm{DRAM}}(S)=\frac{\sum_{e\in S}M_{\mathrm{tmp}}(e)}{M_c}$, and (c) \emph{minimizes} cross-slice total latency $L(\mathcal{E})=\sum_{S\in \mathcal{E}}\max_{e\in S} r_e$. This is a bi-criteria knapsack with a makespan objective and is NP-hard.

Unlike conventional knapsack formulations with one-shot packing and a single objective, our problem is slice-coupled and multi-objective. Greedily maximizing per-slice DRAM can create high-latency “tail” experts, increasing total latency across slices.
To address this, we design a lightweight dynamic programming (DP) algorithm that jointly considers each expert’s temporary DRAM usage and lifespan, while prioritizing lighter experts and biasing toward Shortest-Job-First~\cite{pabla2009completely}. 
This directly optimizes multi-slice latency with maximal DRAM utilization, rather than focusing solely on single-slice throughput or completion time.
Formally, we define each expert's temporary DRAM usage $M_{tmp}^{i}$ as both the item weight $w_{i}$ and value $v_{i} = \frac{M_{tmp}^{i}}{M_{c}} - \eta * r_e$, where $\eta$ is a coefficient to balance these two objectives, allowing for flexibility in different applications.
The dynamic programming algorithm starts from the initial state $dp[0][0]$ (no experts selected, no DRAM used). The state transition equation $dp[i][j] = \max(dp[i-1][j], dp[i-1][j-w_{i}] + v_{i})$ is \textit{iteratively} applied until $dp[N][M_{c}-M{s}]$ yields the optimal expert grouping for $N$ experts under available DRAM. The algorithm iteratively proceeds until all candidate experts in the layer are assigned to ordered time-slices.


\subsubsection{Expert Lifespan-aware Aligning}
\label{sec:resource_all}
After fragmentation-aware grouping, \sysname aligns expert lifespans by allocating GPU resources within each slice to minimize waiting time while preserving accuracy.
The challenge lies in estimating heterogeneous experts’ resource demands (memory- or compute-bound) under dynamic MoE routing and tight, variable edge budgets. Existing learned latency predictors~\cite{wang2023adaevo} or bound-driven profiling~\cite{jia2022codl} perform poorly.
Instead, we pre-profile each expert to record per-token latency and saturation points. 
A work coefficient is computed from token count and saturation, capturing the effective resource need.
The controller then performs a \textit{fill-clamp–refill} allocation: Seed each expert with its minimum required threads; Distribute remaining budget proportional to work coefficients so heavier experts catch up; Clamp saturated experts and redistribute leftover to others.
This iterative procedure quickly balances latency across unsaturated experts, avoids waste beyond saturation, reduces long-tail delays versus uniform splitting, and improves GPU utilization.
In practice, coefficients are refreshed via sliding-window probes with light smoothing, optionally priority-weighted, and integrated with twin-buffer scheduling (\secref{sec:design_3}) to prewarm scheduled experts.

\begin{figure}[t]
    \centering
\includegraphics[width=.46\textwidth]{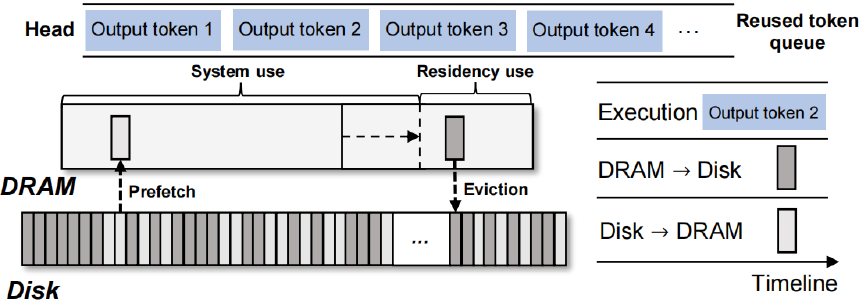}
    \caption{Illustration of token prefetch and eviction.}
    \label{fig:design2}
\end{figure}

\subsection{I/O-efficient Expert Computation Reuse}
\label{sec:design_2}
Co-adaptation gives \sysname a global view of memory and cross-task reuse, enabling I/O planning and cache management that hides external reads behind GPU computation and prioritizes DRAM for high-value items. The goal is to retain hot expert outputs, prefetch next-needed items, and avoid thrashing.
In dynamically sparse V-MoEs, reuse is governed by MoE routing and scenario semantics rather than fixed temporal/spatial locality. Existing prefetch/reuse systems~\cite{guo2018foggycache,guo2018potluck,gao2024attentionstore} assume fixed graphs or steady streams, which fail under dynamic activations: expert firing varies batch to batch, token routing is irregular, and limited DRAM forces frequent evictions. 
Naive prefetching leads to misses, disk$\rightarrow$DRAM stalls, and eviction conflicts, causing latency spikes that nullify reuse gains.

Moreover, as reusable tokens grow, prefetch alone cannot fully overlap I/O with GPU work. 
For example, adapting three V-MoEs with 25\% reusable tokens produces $>50$s I/O latency. 
\sysname addresses this with I/O-hiding token prefetch (\secref{sec:prefetch}) and importance-aware token residency and eviction (\secref{sec:residency}), enabling efficient reuse under large-scale, non-stationary drifts (\figref{fig:our_reuse}).

\subsubsection{I/O Latency-hiding Token Prefetch}
\label{sec:prefetch}
Dynamic sparsity complicates layer-level token reuse, but cross-task token reuse can be \textit{predictable} from expert activation probabilities, as tokens are assigned sequentially in V-MoEs.
\sysname hides I/O latency by overlapping GPU computation with token prefetch: tokens are reordered so non-reusable ones are computed immediately, while reusable tokens are prefetched into DRAM. 
This ensures I/O occurs in parallel with computation, accelerating retraining without blocking GPU work.

\subsubsection{Importance-aware Token Residency and Eviction}
\label{sec:residency}
Existing I/O schedulers and hybrid computation methods interleave tasks but fail under parallel streaming due to isolated expert execution. 
In V-MoE retraining, partial output tokens can be reused across batches, yet they are typically evicted after backpropagation, forcing redundant prefetches.

Intuitively, tokens with high reuse probability and hard-to-hide I/O latency are more valuable and should remain resident in computation-reuse system to maximize I/O latency hiding.
Thus, \sysname proposes importance-aware residency: tokens with high reuse probability and costly-to-hide I/O are retained in DRAM.
%
Difficulty of hiding I/O latency $R_{overlap}^{i}$ for token \(i\) is measured as the ratio of reused tokens \(n_{i}^{e_{reuse}}\) to computed tokens \(n_{i}^{e_{compute}}\) for corresponding expert \(e\) in the current training batch. Higher $R_{overlap}^{i}$ means I/O is heavier relative to compute, and post-compute stalls for I/O latency become more probable, making it more beneficial to keep e’s reuse results resident in DRAM.
Reuse probability $P_{reuse}$ is predicted using a lightweight offline-trained neural network. 
This selectively retains high-value tokens, reducing redundant I/O and improving training efficiency (\figref{fig:our_reuse}),

We collect pixel-level similarity of multiple image pairs and statistical features of experts' outputs (\eg mean, variance, maximum, sparsity) as data to train the model, using reuse success as labels.
Therefore, the importance of output token \(i\) under expert \(e\) in the \(m\)-th training batch is defined as:
$
I_{i}^{e} = R_{overlap}^{i} \times P_{reuse}^{i}= P_{reuse}^{i} \times \frac{n_{i}^{e_{reuse}}}{n_{i}^{e_{compute}}}.
$
For those involving cross-task experts, we extend the importance to:
$
I_{i} = \max_{1 \leq j \leq n} \left\{ P_{reuse}^{i} \times \frac{n_{i}^{j_{reuse}}}{n_{i}^{j_{compute}}} \right\}.
$
After each batch's backpropagation, output tokens with an importance score above a threshold $\lambda$ are retained, while others are evicted.

However, resident output tokens are not always reused in the next batch. 
When successfully reused, they reduce system latency, but if not, they occupy valuable DRAM, adding overhead. 
To manage this, we periodically evict obsolete resident output tokens to avoid unnecessary DRAM consumption, as shown in Fig.~\ref{fig:design2}.
Since one reusable output token can correspond to cross-task experts, eviction risks conflicts. 

To prevent this, we place resident outputs in a dedicated runtime-managed cache region and map them to a contiguous logical buffer space.
Resident output tokens are placed in DRAM from low to high addresses based on the execution order of their last corresponding experts $e^{n}$, ensuring that those more likely to be evicted are positioned at the boundary to avoid fragmentation.
When the resident output token $T_{resident}^{i}$ is reused by expert $e_{i}^{j}$, it is moved to its corresponding sequential address $addr_{e_{i}}^{j}$. After all experts $\{e_{i}^{1}, e_{i}^{2}, ..., e_{i}^{n}\}$ associated with the resident output token $T_{resident}^{i}$ finish execution, it is evicted from DRAM. 
This process repeats for each training batch, maintaining efficient memory management.

\subsection{Asynchronous Resource Scheduling}
\label{sec:design_3}
\sysname decomposes parameter updates into device-specific retraining jobs and schedules them under GPU time-division multiplexing to track per-device adaptation gains. Efficient scheduling is critical under the large-scale, non-stationary drifts. Prior single-buffer schedulers serialize arrivals in only one pipeline, causing head-of-line blocking and high task-switch overhead~\cite{khani2023recl,bhardwaj2022ekya,gu2021liquid}.

To overcome this, \sysname uses a twin-buffer scheduler, \ie asynchronous requests are separated into two buffers, and slices are interleaved across buffers. 
Dynamic estimates (\secref{sec:schduling}) enable sparsity-aware cooperative scheduling, hiding switch costs and converting serialized conflicts into parallel, resource-efficient execution (\secref{sec:twin-buffer}).

\begin{figure}[t]
    \centering
    \includegraphics[width=.45\textwidth]{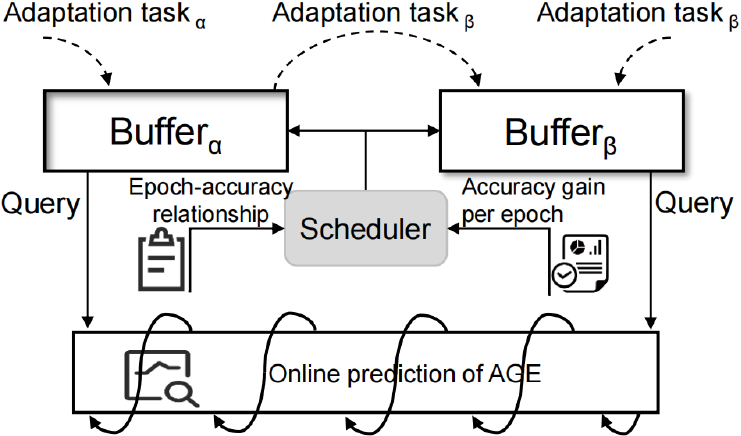}
    \caption{Illustration of Resource Scheduling.}
    \label{fig:design3}
\end{figure}

\subsubsection{Twin-Buffer Scheduler}
\label{sec:twin-buffer}
We observe that
\textit{i)} task requirements reduce to two objectives—avoiding starvation from imbalanced updates and maximizing accuracy within latency constraints;
\textit{ii)} retraining tasks can exploit a two-tier cross-task GPU preemption~\cite{han2024pantheon}, leveraging MoE sparsity to preload non-overlapping experts before task switches, reducing overhead.

Based on this, \sysname introduces a twin-buffer scheduler (\figref{fig:design3}) over black-box assumption, applicable to general asynchronous tasks.
Unlike generic multi-level queues, it exploits V-MoE structure and expert overlap while jointly considering task demands and system operations.
\textit{First, buffer division}
tasks with accuracy and predicted accuracy gain below the lower threshold \(\alpha\) and $\Delta$ are placed in \(buffer_{\alpha}\), while others go into \(buffer_{\beta}\). 
Once a task's accuracy exceeds \(\beta\), it is considered complete and \sysname returns the model. 
\textit{Second, cross-task GPU sharing}. The buffers function as foreground/background pipelines. 
Scheduling addresses task prioritization for GPU cycles, resource allocation per task, and cooperation between buffers to maximize parallelism and minimize switch overhead.
This transforms sparse, asynchronous MoE activations into cooperative scheduling opportunities, improving throughput and lowering latency.

\noindent$\bullet$ For tasks in $buffer_{\alpha}$, the goal is to bring all tasks in large-scale drifts to the performance threshold $\alpha$ quickly, without needing precise latency predictions. 
We apply the \textit{Shortest-Job-First} principle~\cite{elmougy2017novel}, which only requires knowing the time order of tasks reaching $\alpha$, 
and serve tasks for one epoch in a time-slice rotation. 
The \textit{Online prediction of AGE} monitors each task's epoch-accuracy relationship to refine the scheduling order. 
By adjusting based on actual completion time, we prioritize tasks from shortest to longest, minimizing the total time for all tasks to reach $\alpha$. To avoid blocking, we enforce a maximum GPU usage time $\tau$, switching tasks if their accuracy does not reach $\alpha$ within this limit.

\noindent$\bullet$ For tasks in $buffer_{\beta}$, the challenge is maximizing accuracy gains under latency constraints, which is an NP-hard problem. 
Our key insight is that allocating GPU cycles to the task with the fastest accuracy improvement offers an effective heuristic. 
Therefore, we use the \textit{online prediction of AGE} (see~\secref{sec:schduling}) to profile gain per epoch for tasks in $buffer_{\beta}$, then select the task with the fastest accuracy improvement for each GPU cycle until the latency constraints are reached.

\noindent$\bullet$ To facilitate cooperation between the two buffers, we employ an asynchronous pipeline, with \(buffer_{\beta}\) assigned to high-priority streams and \(buffer_{\alpha}\) to low-priority streams. \sysname prioritizes tasks from \(buffer_{\beta}\), and before GPU cycles for \(buffer_{\beta}\) end,
\sysname clears and reloads inactive experts required for the next tasks from \(buffer_{\alpha}\).
When the GPU cycle ends, tasks switch seamlessly, requiring only the change of remaining experts' states and preventing idle GPU time.
Once the next task in \(buffer_{\beta}\) is ready and \(buffer_{\alpha}\) has saved its computation files, \(buffer_{\beta}\) preempts resources for service with the same task switching strategy as above.

\subsubsection{Dynamic Scheduler Inputs}
\label{sec:schduling}
Scheduler requires \emph{accuracy gain per epoch} (AGE) to prioritize slices. 
A key simplification: with a fixed V-MoE architecture and update rule, varying resource schedules (\eg expert parallelism, computation reuse) changes wall-clock time but not the accuracy-\textit{vs.}-epoch trajectory.
\sysname exploits this in epoch space, independent of scheduling decisions. It maintains an online convergence fit of accuracy using a non-negative least squares (NNLS) solver~\cite{nnls}. 
Low-cost early-stopping probes on small data subsets~\cite{standley2020tasks} yield near-term AGE estimates with negligible overhead (< 1ms).
These AGE predictions feed the twin-buffer scheduler, enabling informed foreground and background task prioritization.

\begin{figure}[tbp]
    \centering
    \includegraphics[width=.45\textwidth]{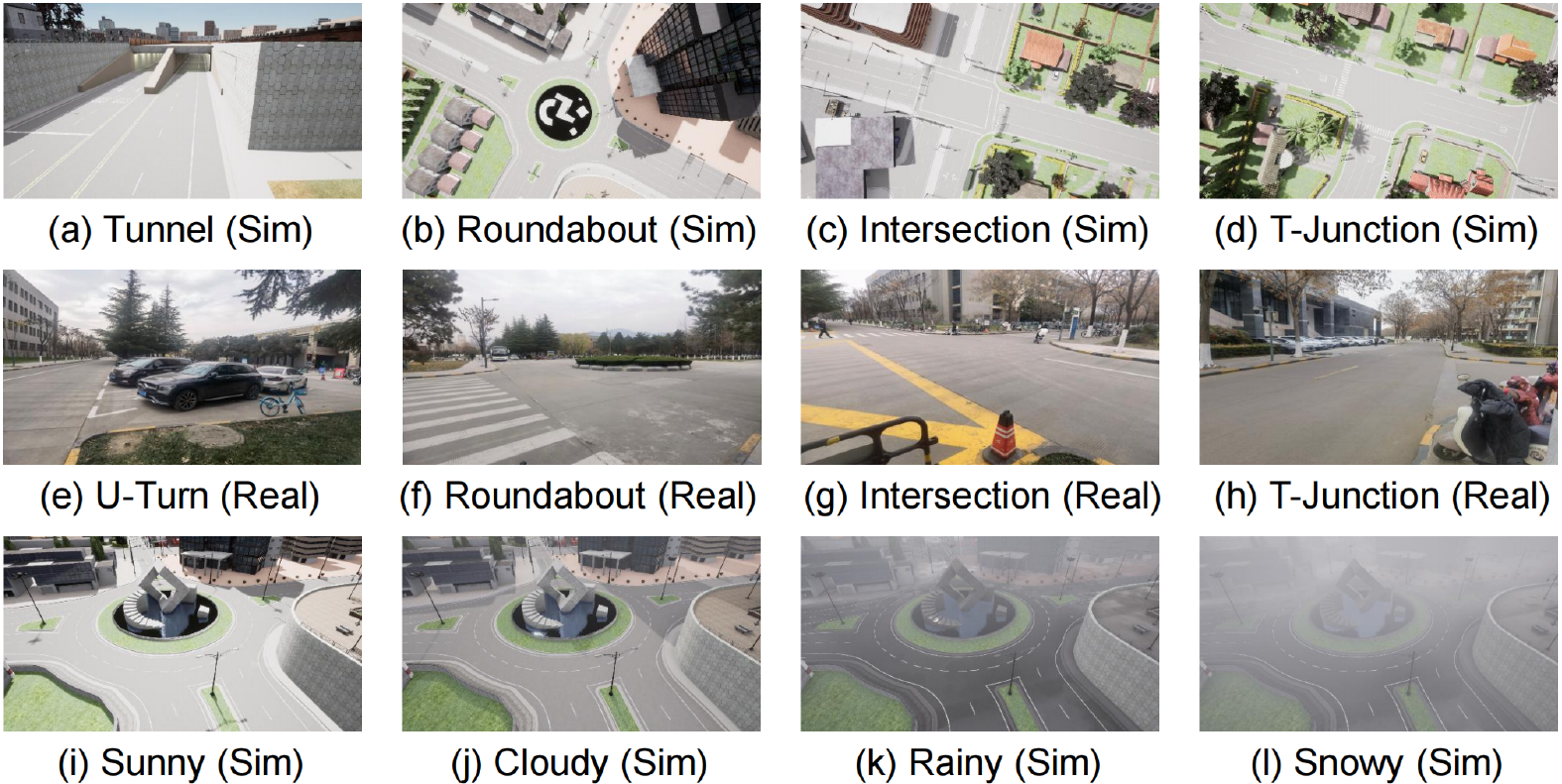}
    \vspace{-1mm}
    \caption{Example scenes in TrafficSim and CampReal.}
    \label{fig:cases}
\end{figure}

\section{System Implementation}
\label{sec:imple}
\sysname adopts a \textit{microservice architecture} for seamless integration with existing V2I frameworks. 
High-performance \texttt{gRPC}~\cite{gRPC} handles inter-service communication, while a lightweight Flask server provides RESTful APIs for management and monitoring. 
\texttt{RabbitMQ}~\cite{RabbitMQ_web} manages asynchronous task queues, decoupling services and improving resilience to workload spikes and moderate network perturbations.

Deployment only requires running the \textit{co-adaptation microservice} on the edge, with devices following a unified model-transfer and data-upload protocol; the existing inference pipeline and V2I stack remain unchanged. 
AdaSprite primarily focuses on optimizing resource utilization at the weak edge. Accordingly, our system is orthogonal to prior work on \textit{adaptation triggers}~\cite{huang2023elastictrainer} and \textit{network optimization}~\cite{meng2024hairpin}, while remaining compatible with them.

\textbf{Parallelism.}
We deploy \sysname under \texttt{NVIDIA MPS}, allowing multiple expert processes to share GPU hardware with minimal context-switch overhead. \texttt{CUDA Streams} overlap data loading, kernel execution, and data movement. Synchronization is enforced only at slice boundaries to match \sysname's slice-level execution.


\textbf{GPU Memory (DRAM) Management.}
We use Unified Memory via \texttt{cudaMallocManaged} to simplify CPU-GPU shared access and overlap transfer with execution. Reusable data are staged in managed buffers and asynchronously prefetched into GPU DRAM via \texttt{cudaMemPrefetchAsync}. 
A custom memory pool recycles frequently used buffers and cache blocks to reduce allocation overhead. The contiguous DRAM region in Sec.~\ref{sec:design} refers to a runtime-managed logical cache layout, rather than physical placement.


\textbf{Asynchronous Performance Monitor.}
We run a lightweight monitor in a separate process using \texttt{multiprocessing}. Through \texttt{py3nvml}, it periodically polls coarse-grained device signals, including available DRAM $(M_{c})$ and aggregate utilization. These signals provide auxiliary feedback for admission control and memory-pressure awareness, while fine-grained scheduling is driven by \sysname's internal runtime state.


\textbf{CARLA Setup.} 
We simulate large-scale V2I traffic drifts using CARLA~\cite{dosovitskiy2017carla} (Town03 map, diverse weather, 150 autopilot vehicles, 38 cameras: 20 on-vehicle, 18 fixed at 2–4m). Synchronous mode captures downsampled frames at 5Hz. 
The code for setup and sampling is available at \url{https://anonymous.4open.science/r/CARLA_setup_sampling/}.

\section{Evaluation}
\label{sec:experiment}

\subsection{Experimental Setup}
\label{exp:setup}
\sysname focuses on computational tasks related to adaptation in vehicle-infrastructure collaboration systems, and is evaluated with reliable communication and data alignment.

\textbf{Dataset and Models.}
We test \sysname on five vision-QA (VQA) tasks in autonomous driving, \ie existence (Exi.), counting (Cou.), query-object (Obj.), query-status (Sta.), and comparison (Com.).
Due to the lack of public VQA datasets for V2I collaboration under large-scale continuous traffic and asynchronous data arrival, we use two self-collected datasets.
\textbf{CampReal}, a 2-week real-world dataset with 1280×720@30FPS videos from six RGB cameras on roadside and AMOVLAB R300 vehicle, capturing traffic in campus from morning to night.
\textbf{TrafficSim}, a CARLA-based simulated dataset with more than 10 road types, 150 vehicles, 38 cameras (vehicle- and roadside-mounted), covering tens of task combinations over 100 hours of continuous traffic, with asynchronous arrivals. 
It spans 4 weather conditions: sunny ($D_{1}$), cloudy ($D_{2}$), rainy ($D_{3}$), and snowy ($D_{4}$).
\figref{fig:cases} shows example scenes from both datasets.



We employ V-MoE as the visual module deployed on devices and distill four compressed versions (V-MoE-85M/-65M/-43M/-22M). 
To perform QA tasks at the edge, we adopt pretrained decoder-only LLMs, including OPT~\cite{opt} with parameter sizes from 1.3B to 2.7B and BLOOM~\cite{BLOOM_web}.
For answer generation, we format prompts as “question:{Q}</s>answer:{A}</s>”, where “</s>” denotes the end-of-sequence (EOS) token provided by the OPT and BLOOM tokenizers and embedded during pretraining.
In addition, we use a well-trained ViT as the teacher model for adaptation and edge performance comparison.

\textbf{Mobile and IoT Devices.}
We use 20 mobile/embedded IoT devices across four platforms: Jetson Nano ($E_{1}$), Jetson Xavier NX ($E_{2}$), Jetson Orin Nano ($E_{3}$), and Jetson AGX Xavier ($E_{4}$), all running V-MoE inference on real-time video streams.
Two edge servers are equipped with two RTX 3080 GPUs (10 GB DRAM each) and one RTX 3090 GPU (24 GB DRAM).




\textbf{Baselines:}
We compare \sysname with four categories of representative adaptation methods:

\noindent$\bullet$ \textbf{No adaptation:} A pre-trained V-MoE without adaptation.

\noindent$\bullet$ \textbf{Isolated on-device model adaptation:}

- Tent~\cite{wang2020tent}: selectively updates normalization layers of V-MoEs through entropy minimization for efficient adaptation.

- EATA~\cite{niu2022efficient}: samples local data and updates part of layers to fine-tune the V-MoE efficiently.

\noindent$\bullet$ \textbf{Edge-assisted isolated model adaptation:}

- Adaptive Model Streaming (AMS)~\cite{khani2021real}: uses a round-robin GPU sharing at edge server across multiple tasks.

- RECL~\cite{khani2023recl}: reuses models in history.

\noindent$\bullet$ \textbf{Edge-assisted model co-adaptation:}

- FoggyCache~\cite{guo2018foggycache}: reuses computation across models without I/O latency-hiding optimization.

- FedAvg~\cite{mcmahan2017communication}: enables collaboration between different task models through model aggregation.

\noindent$\bullet$ \textbf{Edge-assisted MoE-based co-adaptation:}

- MoE-SE: sequentially executes each layer.

- AdaMV-MoE~\cite{chen2023adamv}: balances loads using auxiliary loss.

We measure accuracy using the exact match ratio against the teacher model~\cite{bhardwaj2022ekya,khani2023recl}, and define adaptation latency as the extra time beyond on-device inference for \textit{Tent}, \textit{EATA}, and \textit{FedAvg}.
Additionally, we consider two V2I user-experience metrics:
\textbf{SLO attainment}: fraction of adaptation tasks achieving both $\geq$10\% accuracy gain and $\leq$200s latency;
\textbf{Throughput}: rate of tasks meeting their SLOs.





\begin{table}[tbp]
\centering
\caption{Comparison in accuracy-latency trade-off.}
\vspace{-2mm}
\scalebox{0.65}{
\begin{tabular}{@{}llcccccc@{\hspace{1mm}}c@{}}
\toprule
\multicolumn{2}{c}{\multirow{2.2}{*}{\textbf{Methods \& LLM}}} & \multicolumn{6}{c}{\textbf{Accuracy gain (\%)}} & \multirow{2.2}{*}{\makecell{\textbf{Avg. latency} \\ \textbf{(s)}}} \\ \cline{3-8}
\multicolumn{2}{c}{} & \textbf{Exi.} & \textbf{Cou.} & \textbf{Obj.} & \textbf{Sta.} & \textbf{Com.} & \textbf{Avg.} & \\ \midrule
\multicolumn{1}{c}{\multirow{6.2}{*}{\textbf{OPT-2.7B}}} & Tent~\cite{wang2020tent}      & 3.7  & 21.1 & 29.9 & 10.7 & 8.8  & 14.8 & 394 \\
\multicolumn{1}{c}{}                         & EATA~\cite{niu2022efficient}    & 3.9  & 21.6 & 30.6 & 11.0 & 9.3 & 15.3 & 390 \\
\multicolumn{1}{c}{}                         & FedAvg~\cite{mcmahan2017communication}    & 7.4  & 15.2 & 26.8 & 10.1 & 9.1  & 13.7 & 411 \\
\multicolumn{1}{c}{}                         & AMS~\cite{khani2021real}       & 3.7  & 26.1 & 31.9 & 19.7 & 12.3 & 18.7 & 386 \\
\multicolumn{1}{c}{}                         & RECL~\cite{khani2023recl}      & 3.1  & 41.9 & \textbf{60.7} & 16.8 & 10.1  & 26.5 & 263 \\
\multicolumn{1}{c}{}                         & AdaMV-MoE~\cite{chen2023adamv} & 6.3  & 41.1 & 43.6 & 21.1 & 12.8 & 25.0 & 277 \\
\multicolumn{1}{c}{}                         & AdaSprite & \textbf{12.8} & \textbf{48.3} & 52.5 & \textbf{26.5} & \textbf{18.7} & \textbf{31.8} & \textbf{218} \\ \midrule
\multicolumn{1}{c}{\multirow{6.2}{*}{\textbf{OPT-1.3B}}} & Tent~\cite{wang2020tent}      & 5.4  & 28.2 & 39.5 & 14.8 & 12.7 & 20.1 & 394 \\
\multicolumn{1}{c}{}                         & EATA~\cite{niu2022efficient}    & 5.3  & 29.5 & 40.6 & 14.9 & 13.2  & 20.7 & 390 \\
\multicolumn{1}{c}{}                         & FedAvg~\cite{mcmahan2017communication}    & 10.5 & 25.8 & 37.6 & 14.2 & 13.5 & 20.3 & 411 \\
\multicolumn{1}{c}{}                         & AMS~\cite{khani2021real}       & 5.1  & 31.1 & 41.2 & 27.1 & 16.8 & 24.3 & 331 \\
\multicolumn{1}{c}{}                         & RECL~\cite{khani2023recl}      & 4.8  & 47.8 & \textbf{68.8} & 23.6 & 12.8 & 31.6 & 228 \\
\multicolumn{1}{c}{}                         & AdaMV-MoE~\cite{chen2023adamv} & 9.6  & 46.5 & 54.2 & 29.2 & 18.1 & 31.5 & 236 \\
\multicolumn{1}{c}{}                         & AdaSprite & \textbf{12.4} & \textbf{52.6} & 61.7 & \textbf{34.8} & \textbf{23.3} & \textbf{37.0} & \textbf{193} \\ \midrule
\multicolumn{1}{c}{\multirow{6.2}{*}{\textbf{BLOOM-1B1}}} & Tent~\cite{wang2020tent}      & 3.3  & 17.1 & 22.3 & 7.5  & 6.6  & 11.4 & 394 \\
\multicolumn{1}{c}{}                         & EATA~\cite{niu2022efficient}    & 3.6  & 17.9 & 23.0 & 8.1 & 7.8  & 12.1 & 390 \\
\multicolumn{1}{c}{}                         & FedAvg~\cite{mcmahan2017communication}    & 7.1  & 12.9 & 19.7 & 6.9  & 7.9  & 10.9 & 411 \\
\multicolumn{1}{c}{}                         & AMS~\cite{khani2021real}       & 2.3  & 21.7 & 23.2 & 12.3 & 8.2  & 13.5 & 263 \\
\multicolumn{1}{c}{}                         & RECL~\cite{khani2023recl}      & 2.2  & 35.5 & \textbf{53.8} & 10.1 & 5.0  & 21.3 & 194 \\
\multicolumn{1}{c}{}                         & AdaMV-MoE~\cite{chen2023adamv} & 5.6  & 37.2 & 37.7 & 14.3 & 7.7  & 20.5 & 209 \\
\multicolumn{1}{c}{}                         & AdaSprite & \textbf{10.7} & \textbf{39.4} & 46.2 & \textbf{21.3} & \textbf{14.7} & \textbf{26.5} & \textbf{173} \\
\bottomrule
\end{tabular}}
\label{tab:accuracy_latency_optimized}
\vspace{-4mm}
\end{table}

\subsection{Performance Comparison}


We evaluate \sysname against six baselines, \ie Tent~\cite{wang2020tent}, EATA~\cite{niu2022efficient}, FedAvg~\cite{mcmahan2017communication}, AMS~\cite{khani2021real}, RECL~\cite{khani2023recl}, and AdaMV-MoE~\cite{chen2023adamv}, over three LLMs (\eg OPT-2.7B/1.3B and BLOOM-1B1), focusing on the accuracy-latency trade-off. 
Using five V-MoE models on different devices with 34Mbps video streams in TrafficSim ($D_2$), 
\sysname achieves the best trade-off, improving accuracy by up to 2.4$\times$ and reducing latency by up to 57.9\% through parallel execution and I/O-hiding reuse.
Furthermore, it surpasses SOTA methods, increasing SLO attainment by 1.6$\times$ and throughput by 2.1$\times$ via cross-task co-adaptation and a twin-buffer scheduler that maintains tail accuracy and latency guarantees.




\subsection{Scalability to Task Concurrency}
\label{exp_concurrency}
\begin{figure}[tbp]
    \centering
    \includegraphics[width=.42\textwidth]{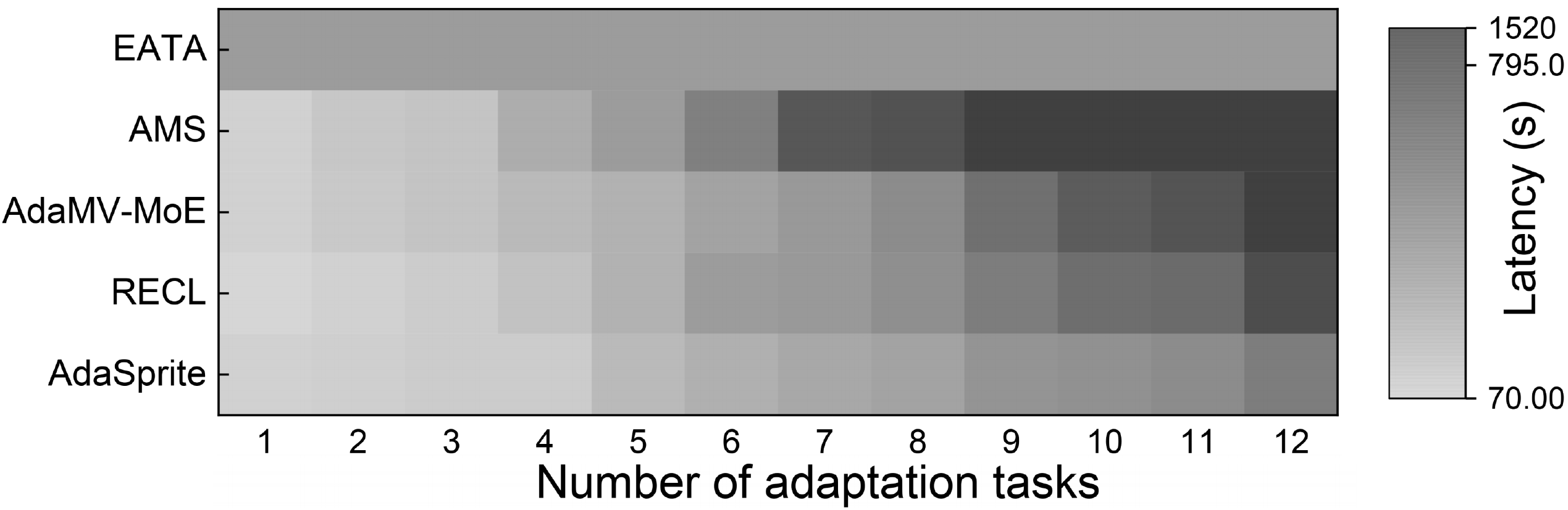}
    \vspace{-0.2cm}
    \caption{Latency comparison across task amounts.}
    \label{fig:latency}
\end{figure}

We evaluate \sysname on 1$\sim$12 concurrent adaptation tasks using OPT-2.7B on TrafficSim ($D_2$), as shown in \figref{fig:latency}.
%
\textit{First}, latency rises with task count for all baselines except \textit{EATA}~\cite{niu2022efficient}. 
\sysname achieves the lowest latency up to 8 tasks, reducing it by up to 56.5\% thanks to DRAM-efficient sparse-expert parallelism and I/O-aware computation reuse, which exploit cross-task redundancy.
\textit{Second}, while isolated adaptations saturate around 6 tasks, \sysname handles up to 17 tasks simultaneously via co-adaptation and efficient off-chip storage. 
\textit{Third}, by tuning accuracy thresholds and maximum GPU rounds ($\tau$), it can trade concurrency and accuracy for lower latency. 
For example, limiting to 4 concurrent tasks with partial data achieves second-level adaptation latency with 15.04\% average accuracy gain.


\begin{figure}[t]
\centering
\begin{minipage}[t]{0.22\textwidth}
\includegraphics[width=3.8cm]{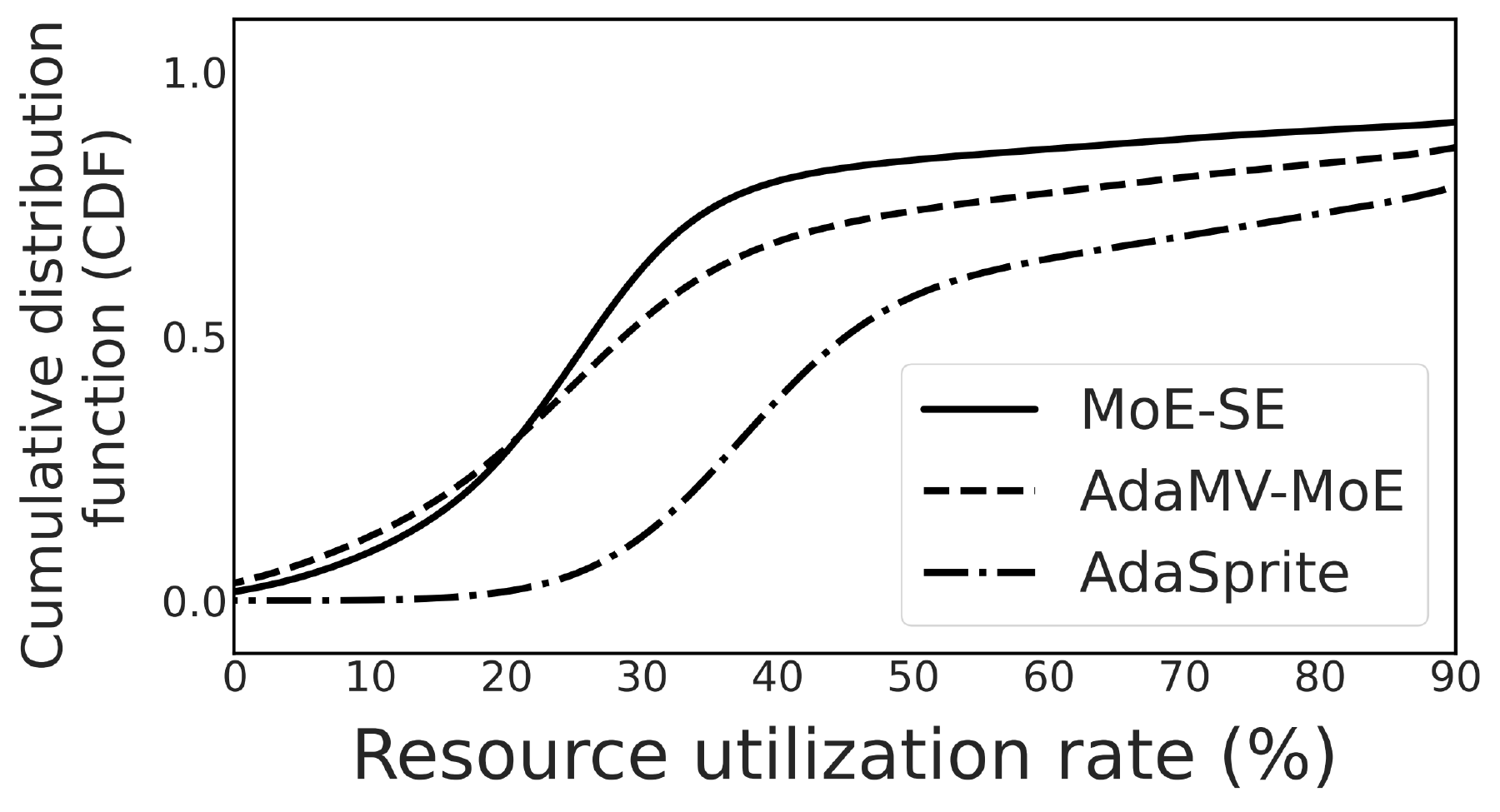}
\centering
            \vspace{-0.2cm}
		\caption{Resource utilization comparison.}
            \vspace{-0.32cm}
		\label{fig:workload}
\end{minipage}
\hspace{0.02\textwidth} 
\begin{minipage}[t]{0.21\textwidth}
\centering
\includegraphics[width=2.8cm]{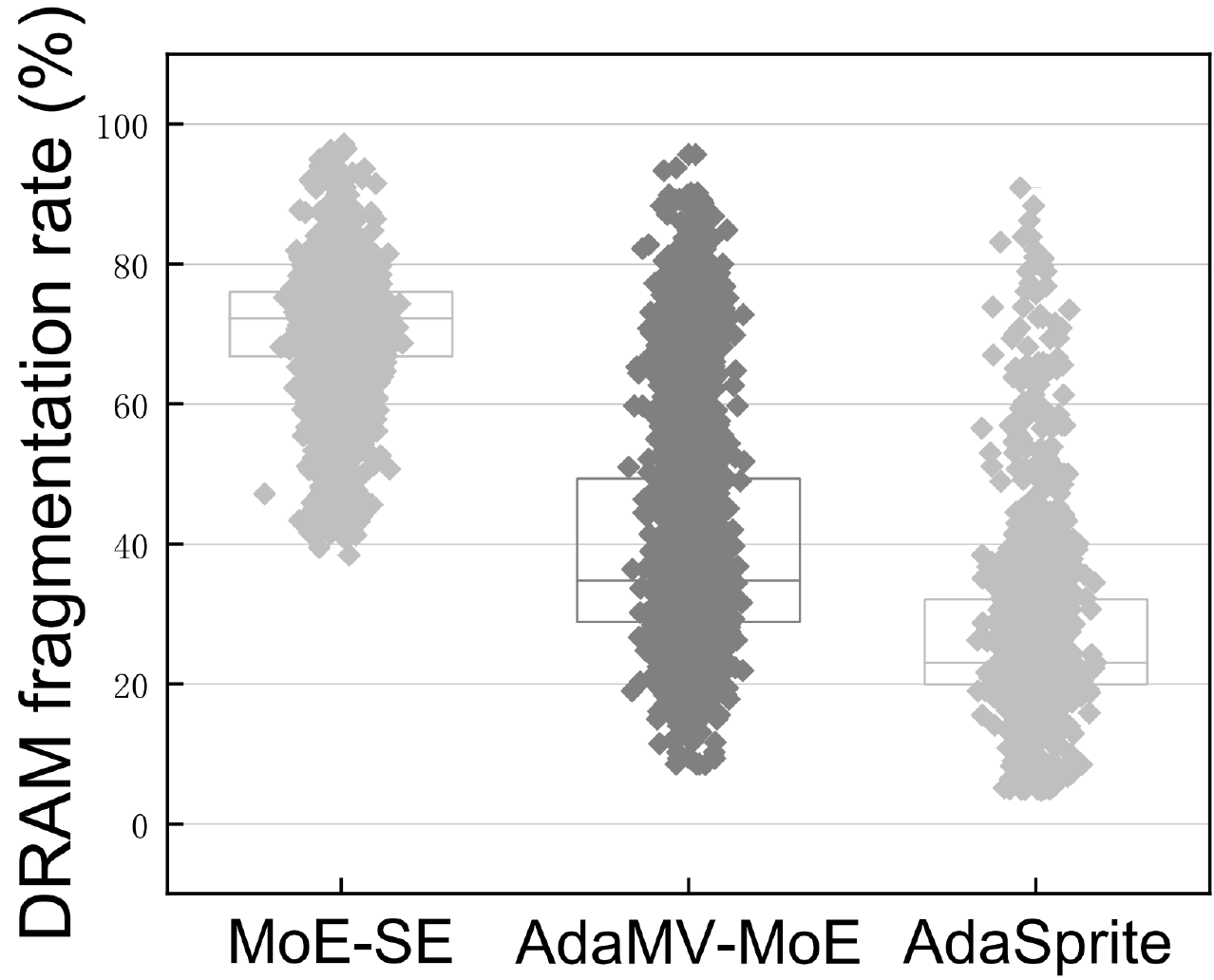}
\centering
            \vspace{-0.2cm}
		\caption{DRAM fragmentation comparison.}
            \vspace{-0.28cm}
		\label{fig:fragment}
\end{minipage}
\vspace{-1mm}
\end{figure}

\subsection{Understanding Improvement}
Each component is evaluated in isolation by replacing it while keeping the rest of \sysname intact.
\label{exp:micro}

\subsubsection{Benefit of DRAM-efficient Sparse Expert Parallelism} 
We compare \sysname with other parallel/serial execution methods using OPT-2.7B on dataset $D_3$.
\figref{fig:workload} shows cumulative GPU utilization, \figref{fig:fragment} compares DRAM fragmentation rates, and \figref{fig:workload_performance} presents accuracy gains.
\textit{First}, \sysname maintains superior GPU utilization (mostly $>$45\% and rarely $<$20\%) by balancing expert loads and reducing stalls.
\textit{Second}, average DRAM fragmentation is reduced by up to 61.9\% via well-aligned execution slices.
\textit{Third}, \sysname outperforms \textit{AdaMV-MoE} in accuracy, as coupling experts across models increases complexity and lowers on-device performance.
\textbf{Overhead:} \textit{Expert grouping and resource allocation add \~1.7 ms per mini-batch.}
\textbf{Extreme case:} \textit{On expert-group failures, fallback to aligned random grouping preserves feasibility while still reducing fragmentation by 42.7\%.}



\subsubsection{Benefit of I/O-efficient Computation Reuse}
\begin{figure}[t]
\centering
\begin{minipage}[t]{0.2\textwidth}
\includegraphics[height=1.7cm]{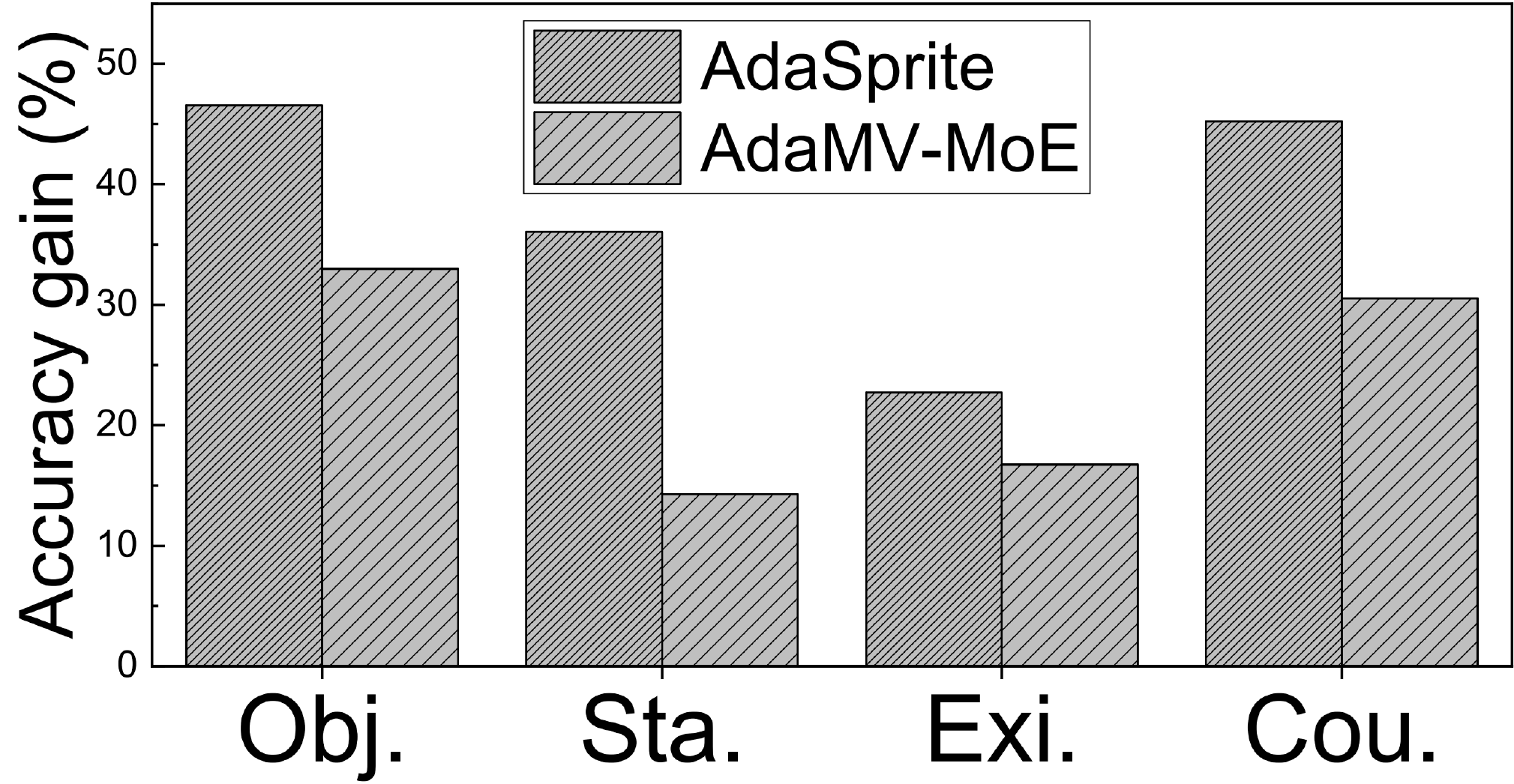}
\centering
            \vspace{-1mm}
		\caption{Comparison of load-balancing.}
		\label{fig:workload_performance}
\end{minipage}
\hspace{0.03\textwidth} 
\begin{minipage}[t]{0.2\textwidth}
\includegraphics[height=1.8cm]{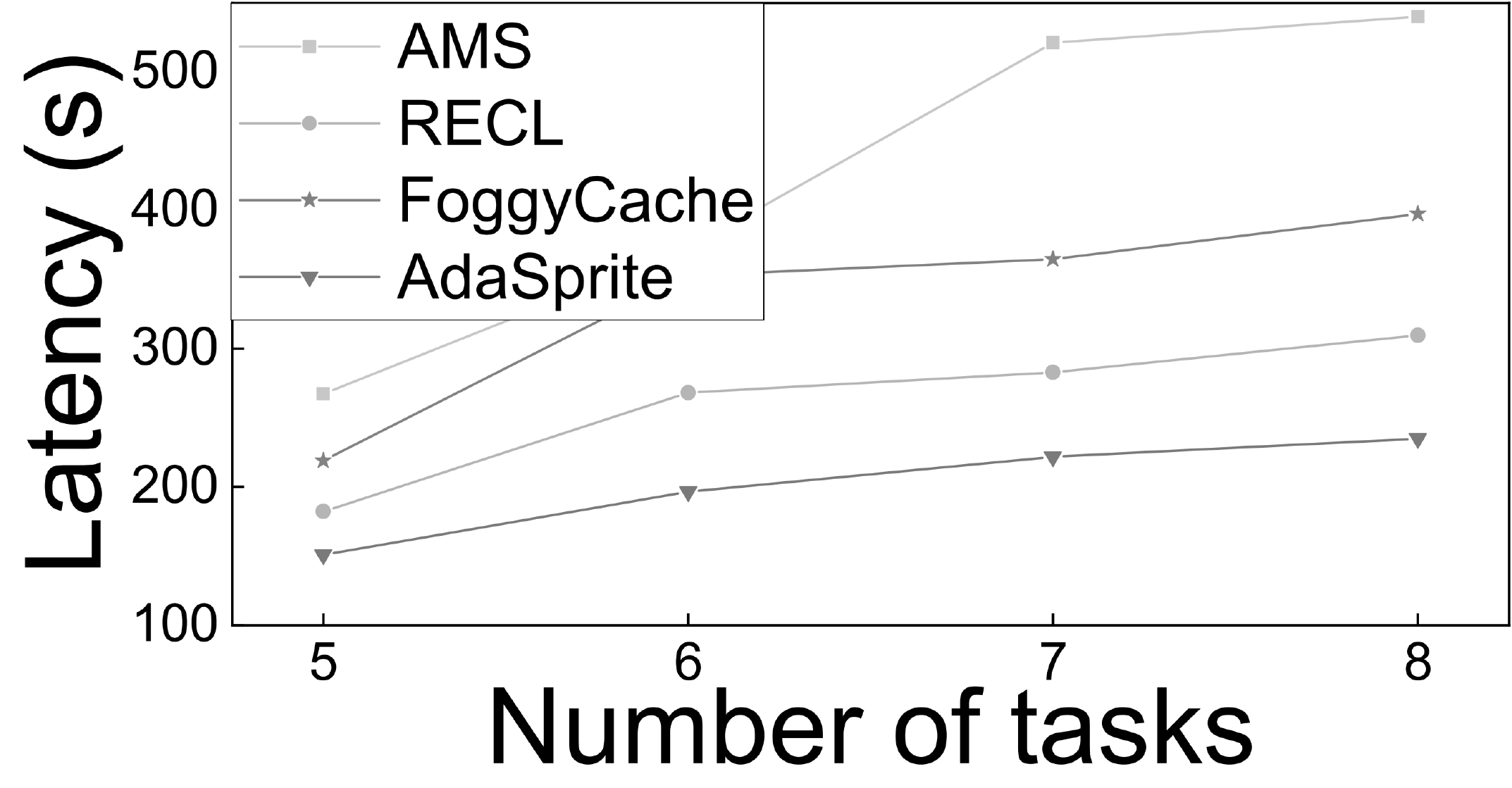}
            \vspace{-1mm}
		\caption{Impact of computation reuse.}
		\label{fig:data_reuse}
\end{minipage}
\vspace{-1mm}
\end{figure}

We compare \sysname with computation reuse methods over $D_{2}$.
Fig.~\ref{fig:data_reuse} shows that \sysname achieves the lowest latency, reducing it up to 40.9\% compared to \textit{FoggyCache}. 
Compared with \textit{FoggyCache}, \sysname cuts I/O wait latency by 86.9\% while reading >30 GB of reusable computation from disk, which is effectively hidden by GPU computing.
\textbf{Overhead:} \textit{I/O-hiding prefetching and importance-aware token residency add around 1.4 ms per mini-batch and use at most 150 MB of GPU memory for buffering.}
\textbf{Extreme case:} \textit{On reuse-prediction errors, \sysname periodically evicts long-idle tokens to prevent blocking.}

%


\begin{figure}[t]
    \centering
    \includegraphics[width=.43\textwidth]{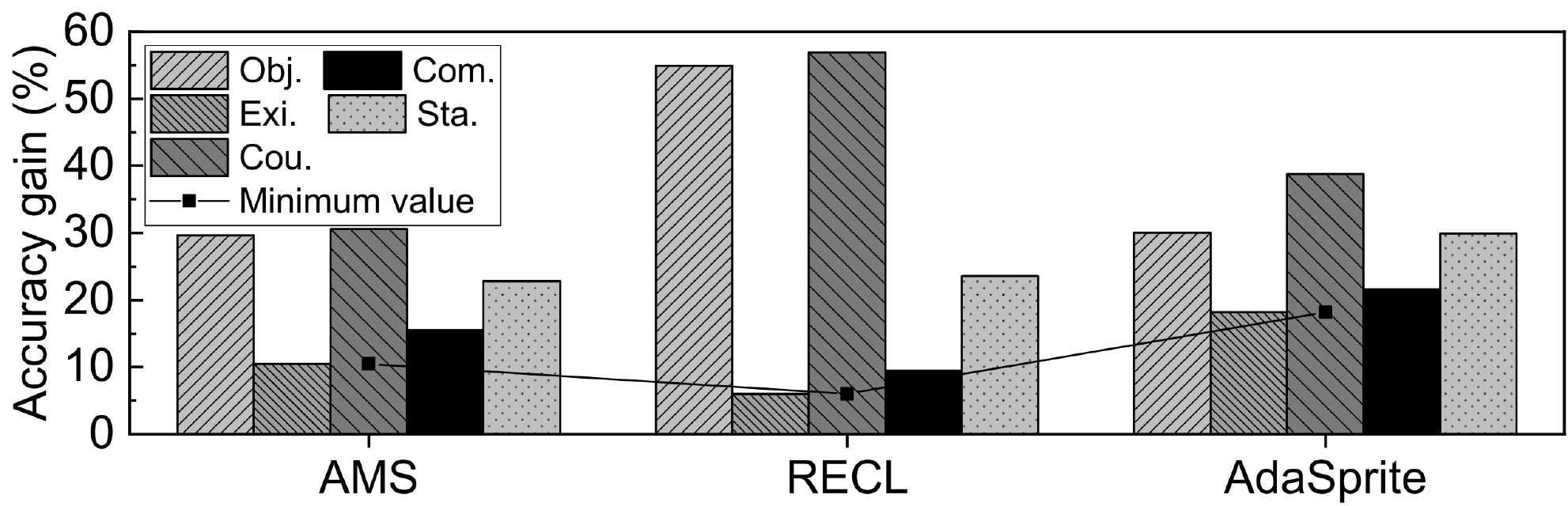}
    \vspace{-2mm}
    \caption{Impact of twin-buffer resource scheduling.}
    \vspace{-1mm}
    \label{fig:schedule}
\end{figure}

\subsubsection{Benefits of Asynchronous Resource Scheduling}
We compare \sysname's accuracy gains with different scheduling strategies, \ie AMS (round-robin) and RECL (time-slice), on TrafficSim ($D_4$) with OPT-2.7B.
As shown in \figref{fig:schedule}, \sysname achieves the best accuracy-stability trade-off, with 27.7\% average and 18.2\% minimum accuracy gains.
Moreover, \sysname boosts SLO attainment by $\geq$2.5$\times$ and throughput by $\geq$3.2$\times$. 
Its twin-buffer scheduler applies buffer-specific policies to maximize accuracy gains, avoid imbalanced updates or starvation, and enforce tail accuracy and latency bounds.
\textbf{Overhead:} \textit{Evaluating 1,000 runs of AGE shows negligible overhead (<1 ms).}
\textbf{Extreme case:} \textit{Facing AGE estimate errors, the scheduler falls back to round-robin and approximate short-job-first behavior. This prevents starvation and long-term resource monopolization, still achieving a 20.8\% accuracy gain.}



\subsection{Micro-benchmark}

\subsubsection{Parameter Sensitivity of Scheduler Threshold $\alpha$ and Residency Threshold $\lambda$}
\begin{figure}[]
\centering
\begin{minipage}[t]{0.2\textwidth}
\includegraphics[width=3.3cm]{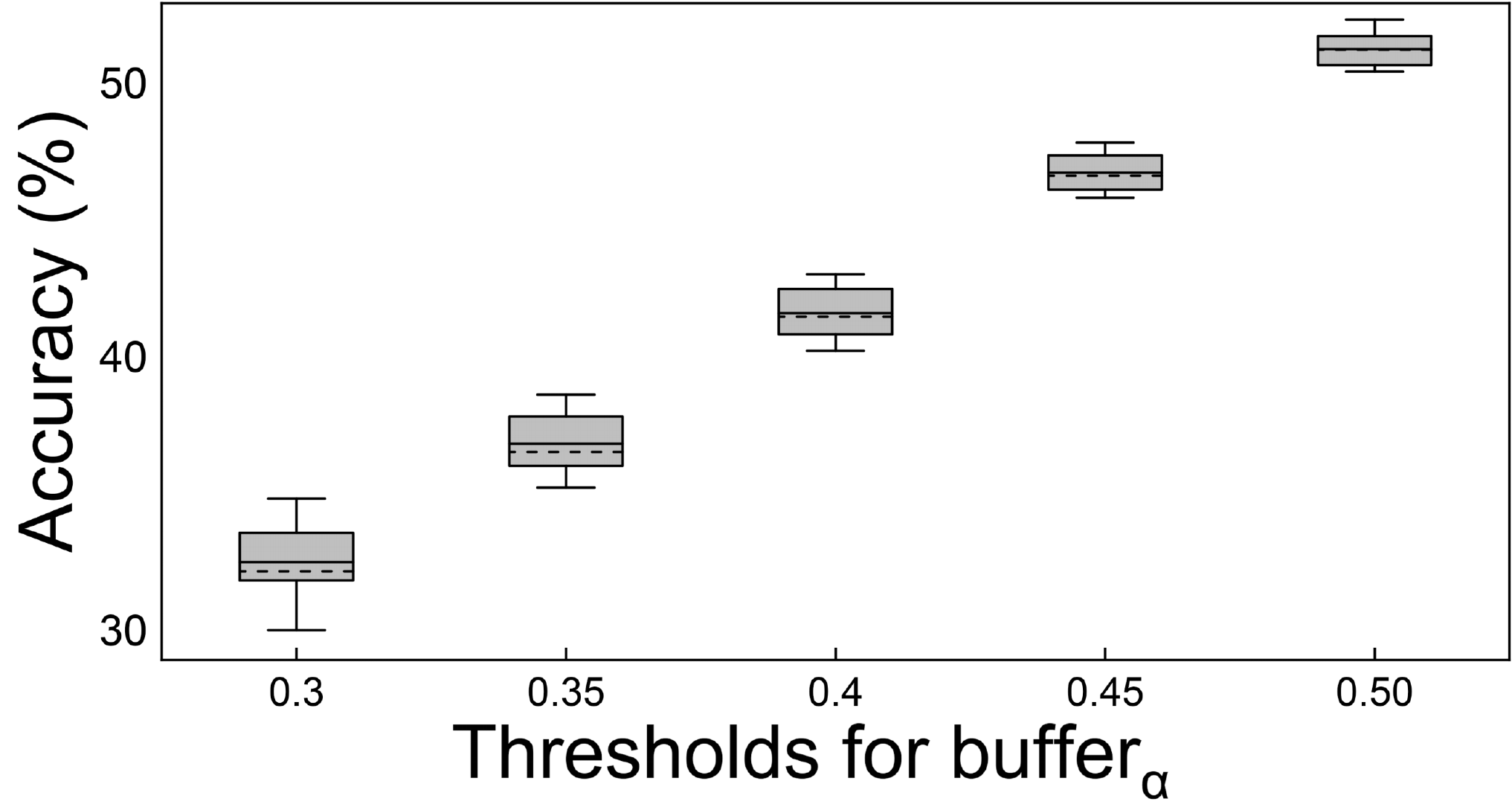}
\centering
            \vspace{-0.1cm}
		\caption{Impact of threshold in scheduler.}
		\label{fig:buffer}
\end{minipage}
\hspace{0.03\textwidth} 
\begin{minipage}[t]{0.2\textwidth}
\includegraphics[width=3.2cm]{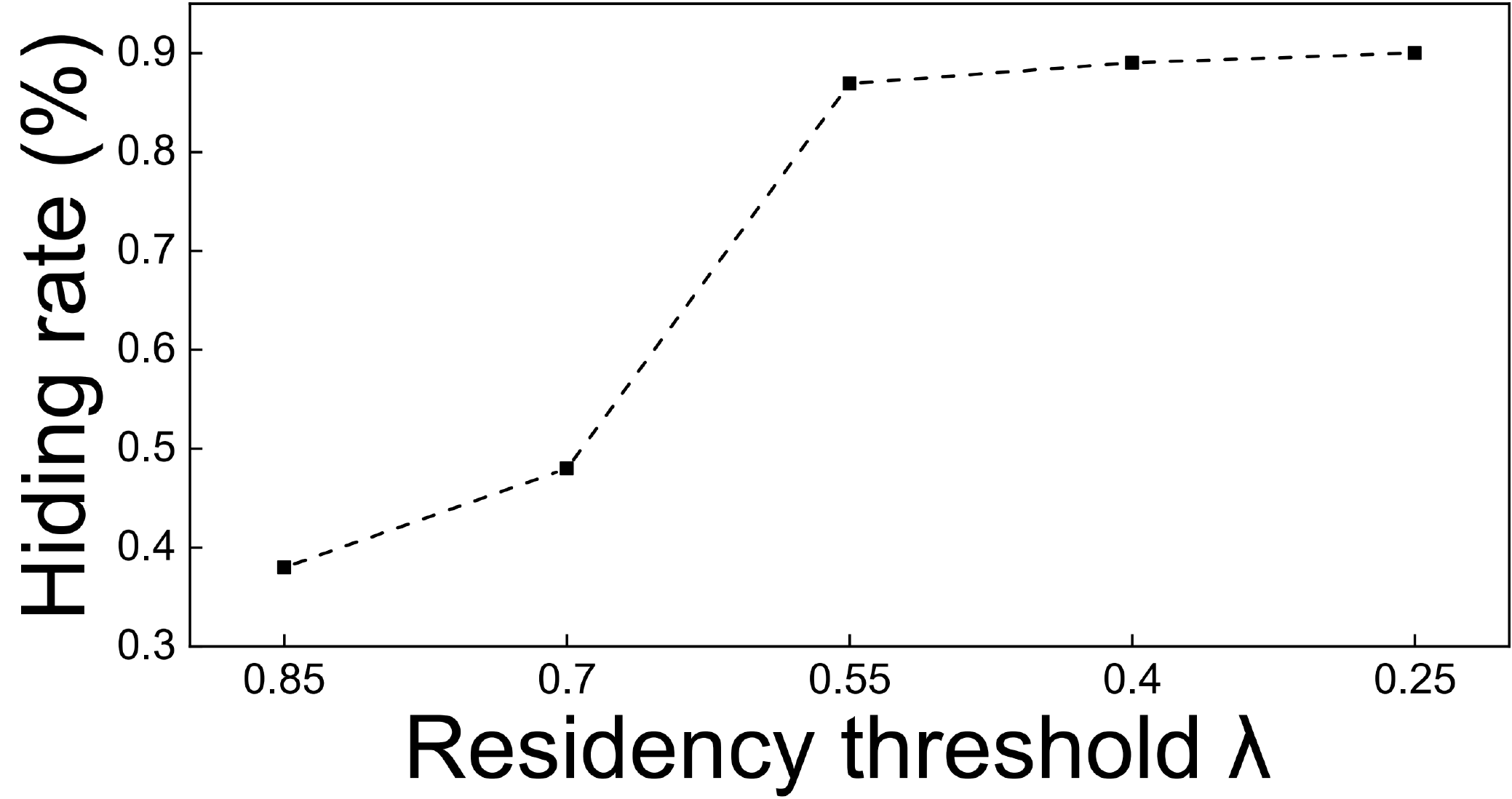}
            \vspace{-0.1cm}
		\caption{Impact of residency threshold.}
		\label{fig:residency}
\end{minipage}
\vspace{-2mm}
\end{figure}

Across four TrafficSim scenarios, \figref{fig:buffer} and~\ref{fig:residency} show average performance under varying thresholds. 
\sysname is generally robust to them.
For example, as for Exi., sweeping $\alpha$ from 0.3$\sim$0.5 shows that smaller values reduce mean accuracy, while larger values improve tail accuracy with higher latency. 
The choice of $\alpha$ can be tuned based on application requirements.
For $\lambda$, values above 0.55 saturate the hiding rate, but larger thresholds can increase evictions. 
Therefore, we set $\lambda = 0.55$ as a balanced default.



\subsubsection{Performance across Diverse Network Bandwidths}
\begin{figure}[]
  \subfloat[Network latency]{\label{fig:network1}
  \includegraphics[width=0.2\textwidth]{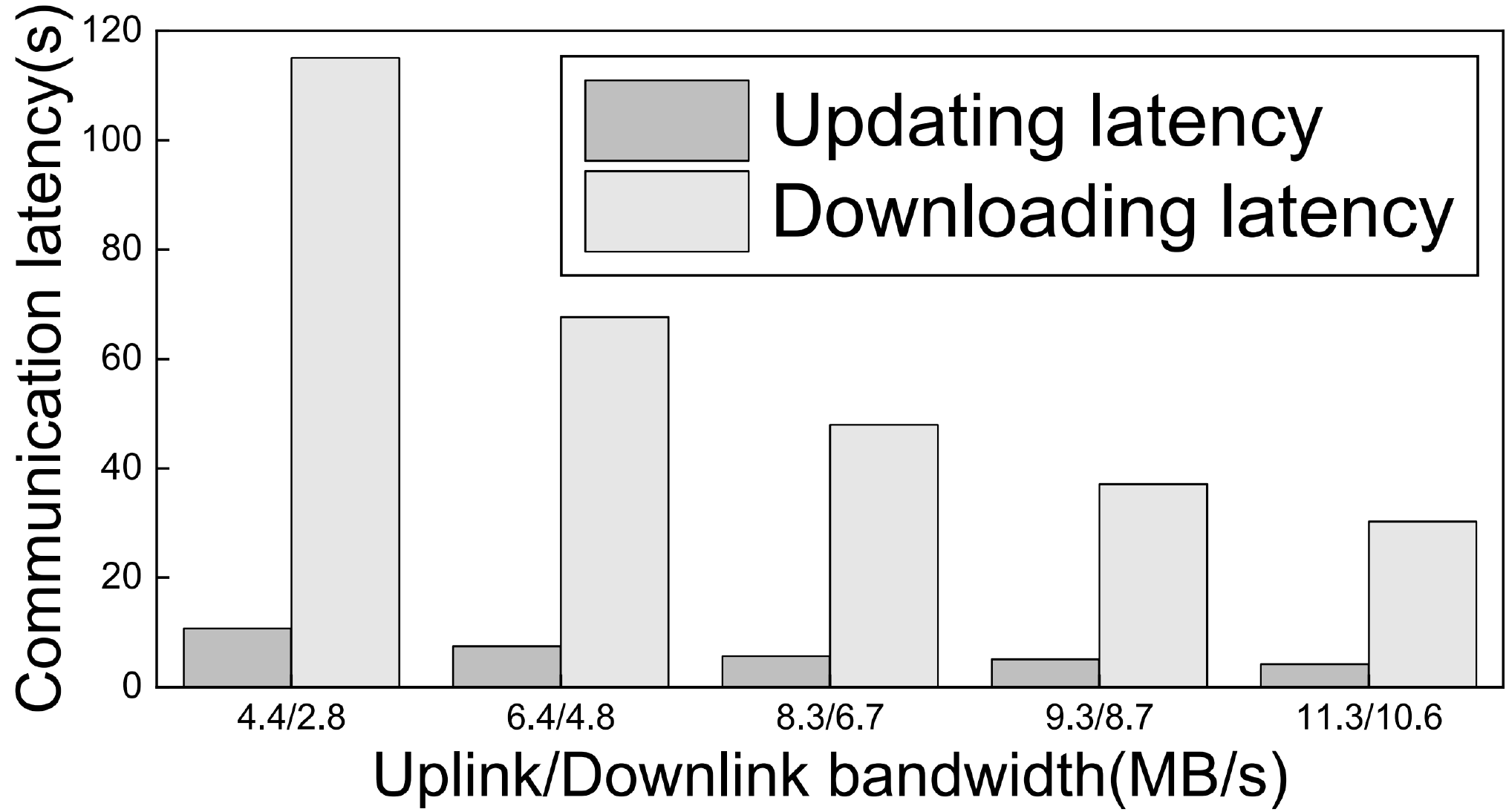}}
  \hfill
    \subfloat[Average accuracy]{\label{fig:network2}
  \includegraphics[width=0.2\textwidth]{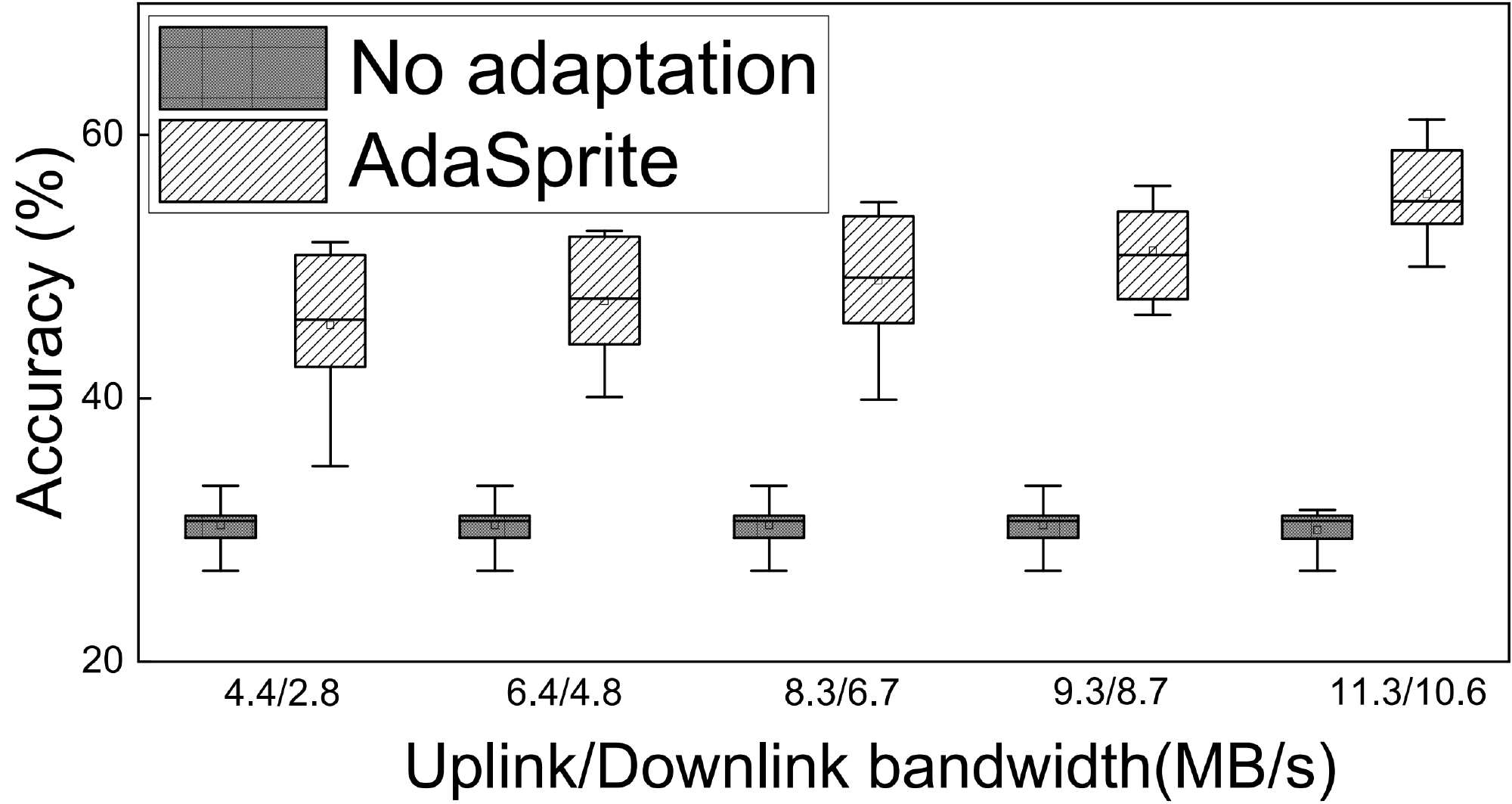}}
  \hfill
  \vspace{-0.1cm}
  \caption{Generalization of network bandwidths.}
  \vspace{-0.1cm}
  \label{fig:network}
\end{figure}

We test \sysname across five network bandwidths using 10 Obj. tasks on TrafficSim $D_4$ (see Fig.~\ref{fig:network}) to assess its robustness under unstable network conditions.
As bandwidth decreases, network latency increases, resulting in higher latency and lower average accuracy. 
Nonetheless, \sysname consistently enhances accuracy by at least 50.1\%, demonstrating generalization across varying network conditions.
%



\begin{figure}[]
  \subfloat[Generalizing to various V-MoEs]
  {\label{fig:vit}
  \includegraphics[width=0.19\textwidth]{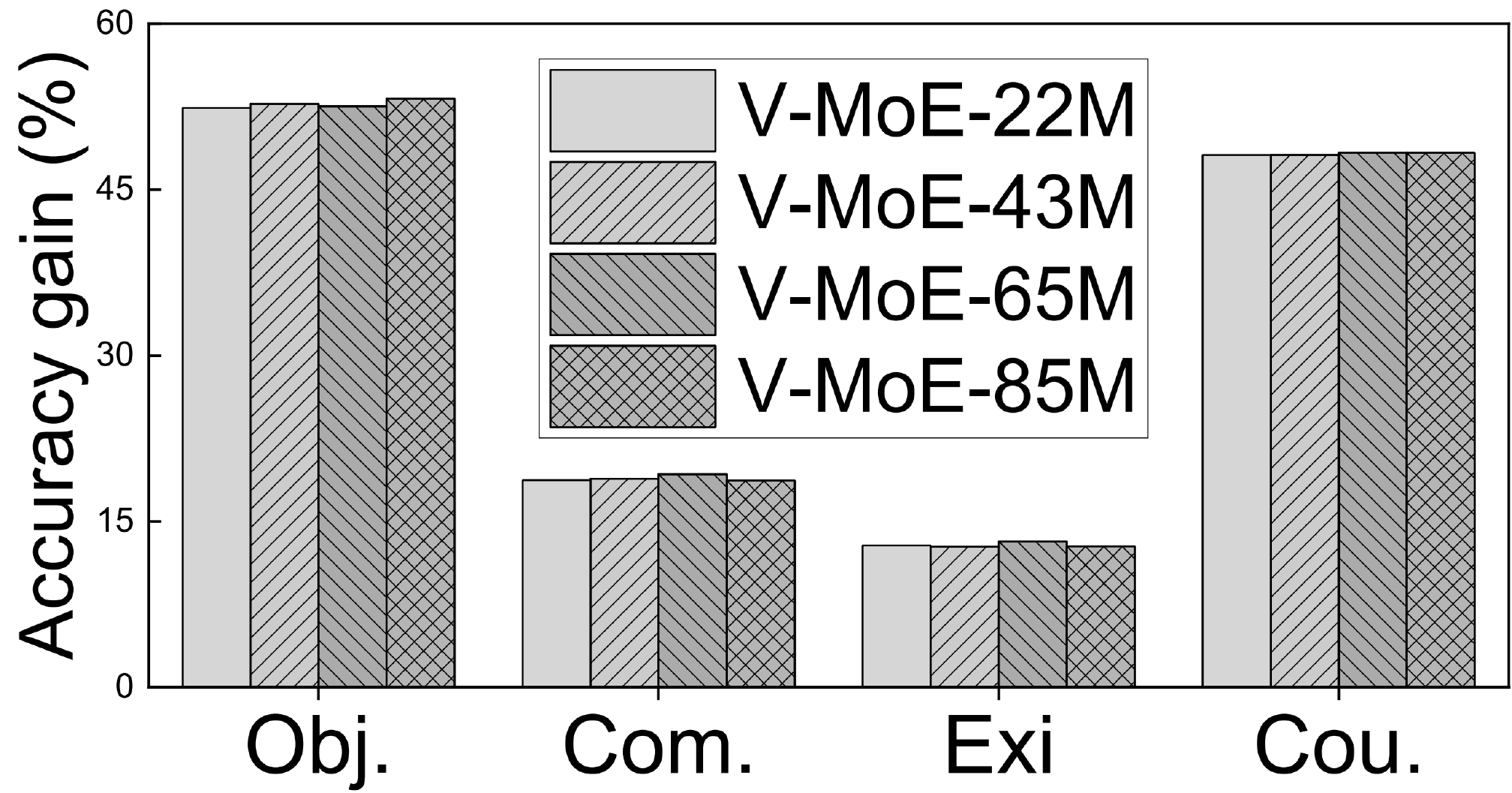}}
  \hfill
    \subfloat[AGE prediction performance]{\label{fig:speed}
  \includegraphics[width=0.19\textwidth]{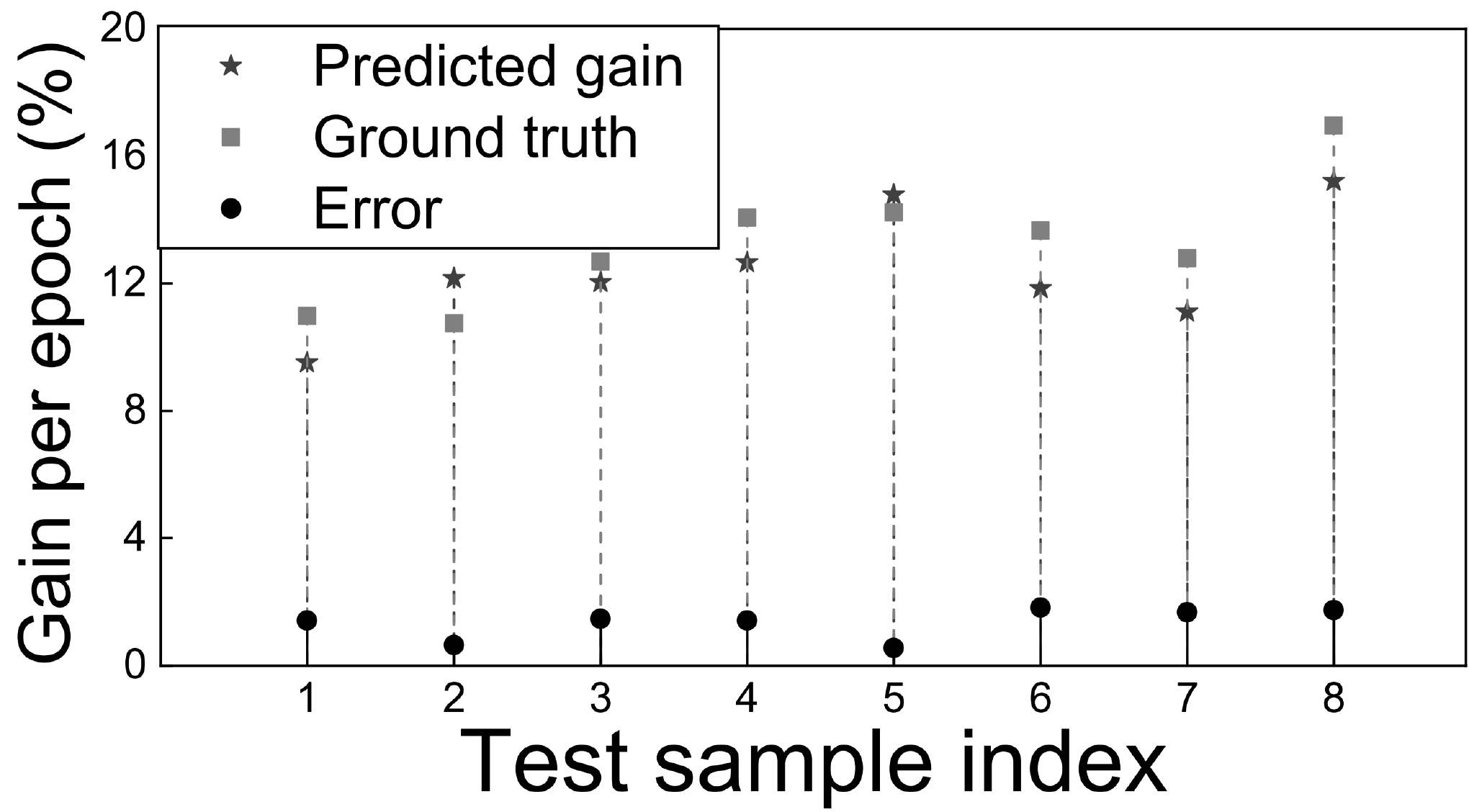}}
  \hfill
  \vspace{-2mm}
  \caption{Generalization of AdaSprite.}
  \vspace{-1mm}
  \label{fig:gernal}
\end{figure}

\subsubsection{Generalizing to Various V-MoEs}
We demonstrate \sysname's ability to robustly calibrate V-MoEs with different sizes deployed on IoT devices ($E_{1}$, $E_{2}$, $E_{3}$, $E_{4}$). 
We evaluate 4 V-MoEs with TrafficSim ($D_{2}$), shown in Fig.~\ref{fig:vit}. 
All V-MoEs achieve significant accuracy gains with \sysname, with at least a 33\% average improvement.


\subsubsection{Performance of AGE Prediction Method}
We evaluate the accuracy gain per epoch (AGE) prediction method from \secref{sec:design_3}. 
Using 8 video segments from $D_{4}$, we test the errors of predicted AGE for Obj., as shown in Fig.~\ref{fig:speed}. 
The error rates remain below 13.3\%, which is acceptable.




\subsection{Case Study}
\label{sec:case_study}
\begin{figure}[]
\vspace{-0.2cm}
  \subfloat[Simulator]{\label{fig:sim_de}
  \includegraphics[width=0.24\textwidth]{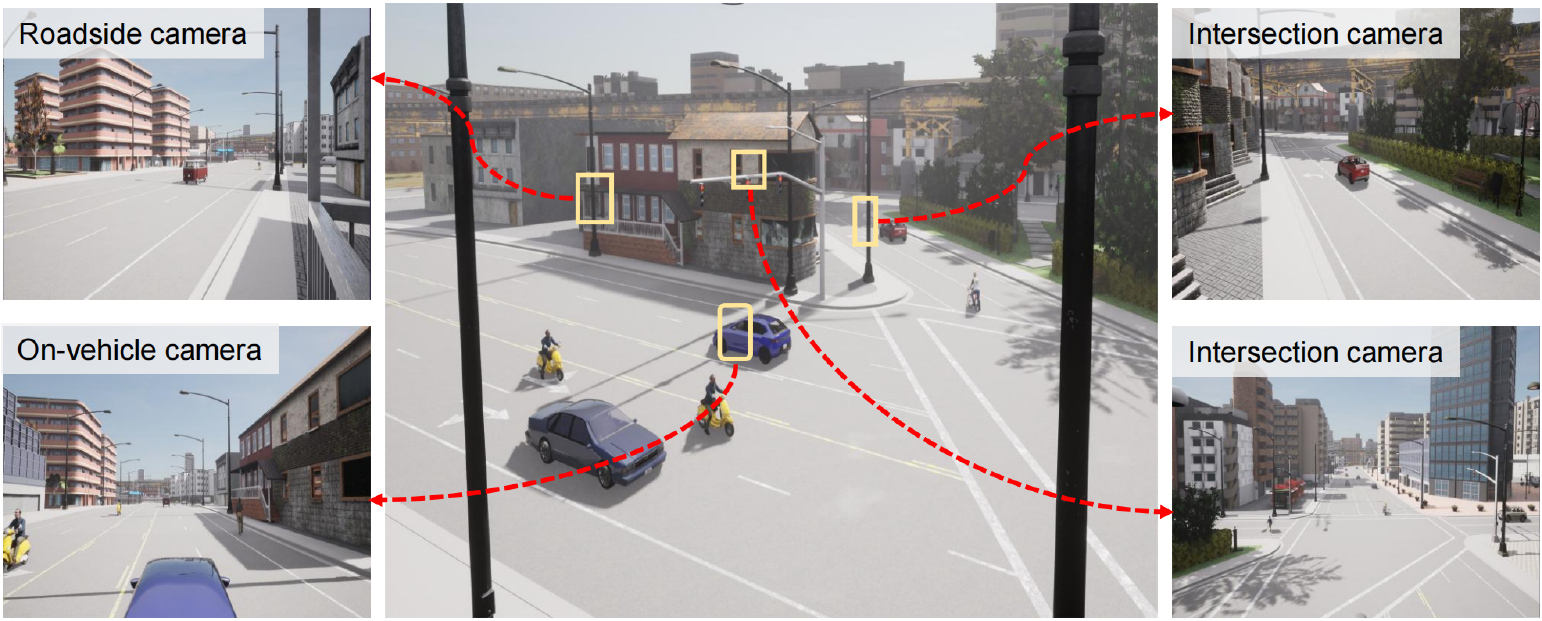}}
  \hfill
    \subfloat[Real data collection]{\label{fig:real_de}
  \includegraphics[width=0.21\textwidth]{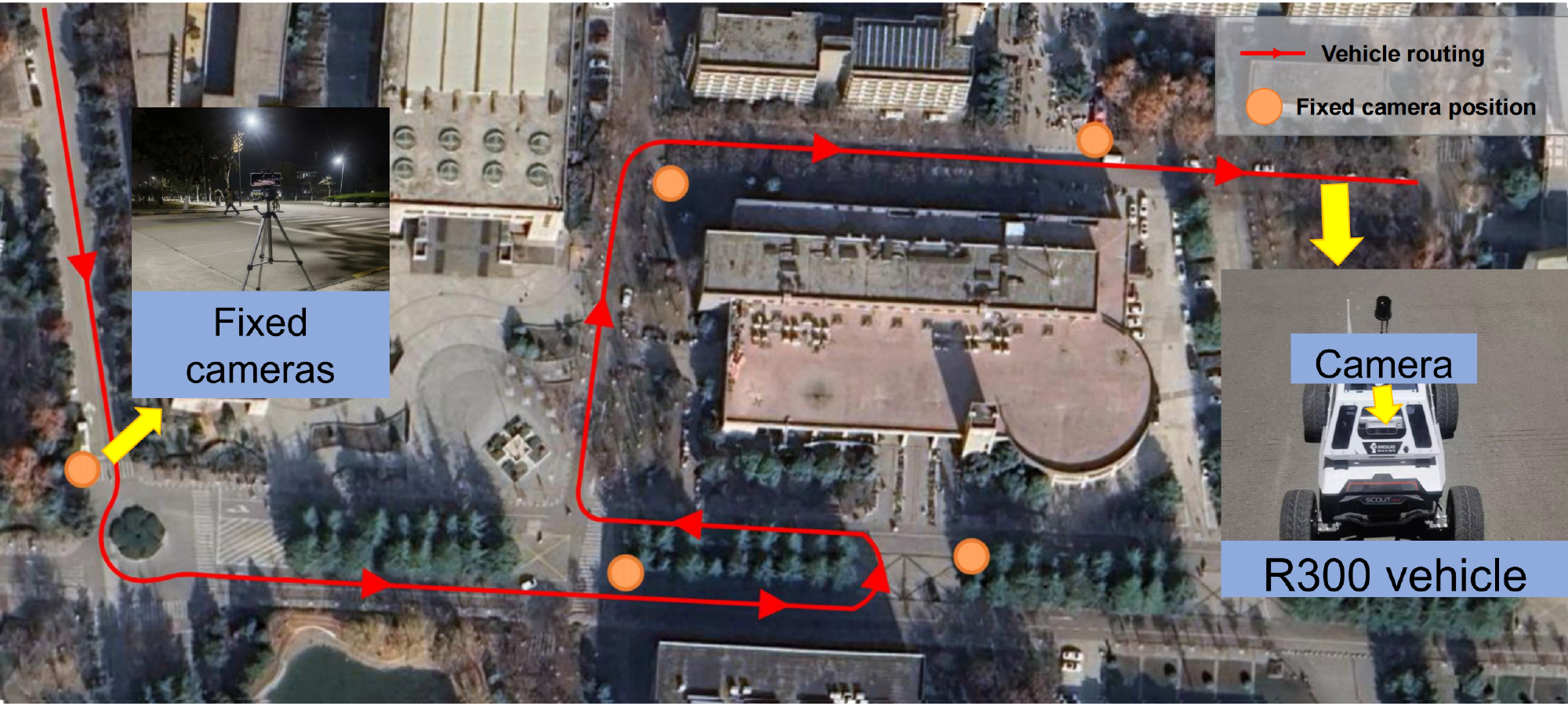}}
  \hfill
  \vspace{-0.2cm}
  \caption{Illustration of case study.}
  \vspace{-0.2cm}
  \label{fig:case}
\end{figure}

We test \sysname in an urban VLM-based (OPT-2.7B) V2I application, comparing it with five baselines using a hybrid simulator and real-world setup.
In simulator, we model urban traffic ($D_{1}$) in CARLA, covering tens of concurrent model combinations and 10+ road types, with 20 on-vehicle cameras and 18 roadside cameras (\figref{fig:sim_de}).
We further validate generalization on the two-week real day-by-day data  (\figref{fig:real_de}).

As shown in Table~\ref{tab:case_study}, \sysname achieves the highest accuracy across four tasks, yielding overall gains of 30.0\% (simulation) and 36.1\% (real-world) over \textit{No adaptation}, demonstrating robust performance in dynamic mobile environments.
While \textit{RECL} performs well on the Obj. task, its accuracy-focused scheduler induces starvation in other tasks, causing severe imbalance. 
\sysname consistently outperforms \textit{RECL} across all other tasks, boosting SLO attainment ($\geq$10\% accuracy gain and <350s latency) by 3.1$\times$ and throughput by 4.3$\times$.
There are two key factors:
\textit{i)} Co-adaptation mechanism optimizes V-MoE convergence and task concurrency, enhancing accuracy gain under limited latency at resource-constrained edge.
\textit{ii)} Parallel optimization and I/O-hiding computation reuse exploit V-MoE’s dynamic sparsity, reducing adaptation latency and runtime even for low-accuracy models.

\begin{table}[]
\caption{Accuracy comparison in continuous traffic.}
\vspace{-2mm}
\scalebox{0.52}{
\begin{tabular}{cccccccc}
\hline
\multirow{3}{*}{\textbf{Metric}}           & \multirow{3}{*}{\textbf{Tasks}} & \multicolumn{6}{c}{\textbf{Methods}}                                                                                                                                     \\ \cline{3-8} 
                                  &                        & \multirow{2}{*}{No \textbf{adaptation}} & \multirow{2}{*}{\textbf{Tent}} & \multirow{2}{*}{\textbf{AMS}} & \multirow{2}{*}{\textbf{RECL}} & \multirow{2}{*}{\textbf{AdaMV-MoE}} & \multirow{2}{*}{\textbf{AdaSprite}} \\
                                  &                        &                                &                       &                      &                       &                            &                            \\ \hline
\multirow{5}{*}{\shortstack{\textbf{Average}\\\textbf{accuracy (\%)}}} & \textbf{Exi. (Sim/Real)}                   & 39.5/32.8                           & 41.5/33.8                  & 42.1/34.7                 & 40.3/33.4                  & 44.4/35.6                       & \textbf{50.2/40.1}                       \\
                                  & \textbf{Cou. (Sim/Real)}                   & 14.2/9.1                           & 15.1/10.3                  & 16.6/11.9                 & 14.6/9.9                  & 15.2/11.4                       & \textbf{20.4/15.9}                       \\
                                  & \textbf{Obj. (Sim/Real)}                   & 20.5/13.2                           & 23.1/15.6                  & 26.6/17.7                 & \textbf{33.2/23.1}                  & 27.4/18.6                       & 31.1/23.0                       \\
                                  & \textbf{Sta. (Sim/Real)}                   & 33.7/25.6                           & 35.7/27.8                  & 39.9/31.7                 & 44.2/34.7                  & 41.6/33.6                       & \textbf{44.7/36.7}                       \\
                                  & \textbf{Com. (Sim/Real)}                   & 61.3/49.5                           & 63.7/52.0                  & 64.3/53.1                 & 63.4/51.8                  & 66.9/52.7                       & \textbf{72.9/61.2}                       \\ \hline
\end{tabular}}
\label{tab:case_study}
\end{table}

\section{Related Work}
\label{sec:related}

\textbf{Online Adaptation Methods.}
Online adaptation addresses data shifts in real-time vision systems. 
Existing methods include \textit{on-device}~\cite{huang2023elastictrainer,niu2024test,niu2022efficient} and \textit{server-assisted} adaptation~\cite{bhardwaj2022ekya,khani2023recl,yang2023edgefm,wang2023adaevo}. 
\textit{On-device adaptation} fine-tunes DNNs using local data during inference. 
However, the high adaptation complexity (\eg a ViT-Base requires over 1.2 GB of peak memory) prevents adaptation on inference-oriented hardware.
%
Additionally, they focus on isolated adaptation, limiting collaboration. 
While federated learning helps~\cite{rajib2025fedctta}, object heterogeneity in V2I can result in accuracy loss~\cite{niu2024test,kirkpatrick2017overcoming}.

\textit{Server-assisted adaptation} leverages edge/cloud resources, allowing mobile/IoT devices to prioritize inference without interference. 
Research primarily focuses on maximizing accuracy under latency constraints~\cite{bhardwaj2022ekya,khani2023recl,kong2023edge}, using techniques like \textit{data resampling}~\cite{khani2021real,mullapudi2019online}, \textit{adaptive triggers}~\cite{kong2023edge,yang2023edgefm}, \textit{sparse updating}~\cite{wang2023adaevo}, and latency optimization with task scheduling~\cite{khani2023recl} and resource allocation~\cite{bhardwaj2022ekya}.
However, most methods are task-isolated and collaboration across tasks is less explored.
In contrast, \sysname uses edge server-assisted adaptation to enable cross-device V-MoE co-adaptation for VLM-based system, ensuring resource efficiency.

\textbf{Edge-assisted Multi-Task Learning.}
Co-adaptation is similar in form and implementation to the widely-researched multi-task learning.
Multi-task learning aiming to construct a unified model improves data and resource efficiency through parameter sharing and knowledge integration.
It has seen applications in edge computing, such as blood glucose modeling~\cite{gu2017sugarmate} and depression detection~\cite{lu2018joint}.
Existing methods focus on modeling task relationships or optimizing joint objectives to maximize benefits, using relationship modeling~\cite{zhang2019pattern,standley2020tasks}, adaptive task learning rates~\cite{chen2018gradnorm}, dynamic parameter sharing~\cite{misra2016cross,sun2020adashare}, and task gradient balancing~\cite{kendall2018multi,yu2020gradient}.
Unlike them, \sysname does not aim to obtain a unified model handling multiple tasks simultaneously, but to efficiently fine-tune V-MoEs separately deployed on devices.

\textbf{Efficient ViT Fine-tuning Techniques}.
Efficient fine-tuning methods for ViT-related models include \textit{additional module tuning}~\cite{yang2023aim,sharma2023lossless}, \textit{prompt-based}~\cite{tsai2023convolutional,wang2024lion,huang2023diversity}, and \textit{specification-based tuning}~\cite{hu2021lora,basu2024strong}.
\textit{Additional module tuning} injects small neural modules or parallel networks into layers, tuning only them for adaptation. 
However, this increases ViT complexity~\cite{hu2021lora}.
\textit{Prompt-based tuning} introduces learnable parameters~\cite{lester2021power} or prefixes~\cite{li2021prefix} into embedding space or input data, which causes slow convergence~\cite{ding2022delta} with non-monotonic performance~\cite{hu2021lora}.
\textit{Specification-based tuning} selectively updates relevant parameters based on specific data/tasks, keeping others frozen. 
We follow this to allow efficient fine-tuning while maintaining accuracy.


\section{Conclusion}
\label{sec:conclude}
In this paper, we present \sysname, a resource-efficient V-MoE co-adaptation system for large-scale VLM-based V2I perception.
By leveraging expert lifespan alignment, I/O-hiding computation reuse, and a twin-buffer scheduler, \sysname turns the challenges of isolated adaptation into co-adaptation opportunities.
Evaluations across diverse scenarios show substantial accuracy and latency gains over state-of-the-art methods, demonstrating a robust and scalable solution for large-scale mobile applications.
Although in V2I applications, real-time inference and information broadcasting enabled by cooperative perception can help address emergencies, thereby complementing adaptation, other applications struggle to handle such situations.
While \sysname allows accuracy-concurrency trade-offs for second-level adaptation, communication overhead still limits real-time operation. 
Rapid cross-device knowledge sharing offers a promising avenue for complementing slower retraining, which we plan to explore in future work.

\section*{Acknowledgments}
We sincerely thank our anonymous shepherd and all reviewers for their valuable feedback. This work is supported by National Key R\&D Program of China (No. 2024YFB4505502), Young Scientists Fund (Category B) of the National Natural Science Foundation of China (No. 62522215), and National Natural Science Foundation of China (No. 62472354).

\bibliographystyle{ACM-Reference-Format}
\bibliography{acmart}

\end{document}